\documentclass[useAMS,usenatbib,usegraphicx]{mn2e}

\usepackage{times}

\newcommand\kms{$\rm km\,s^{-1}$}

\newcommand{\Oiii}{[{\sc O$\,$iii}]}
\newcommand{\Oii}{[{\sc O$\,$ii}]}

\newcommand{\Ha}{H$\alpha$}
\newcommand{\Hb}{H$\beta$}
\newcommand{\Hd}{H$\delta$}
\newcommand{\HdA}{H$\delta$A}
\newcommand{\Hi}{{\sc H$\,$i}}
\newcommand{\Hii}{{\sc H$\,$ii}}

\newcommand{\Nii}{[{\sc N$\,$ii}]}
\newcommand{\Sii}{[{\sc S$\,$ii}]}

\newcommand{\HST}{{\it HST\/}}

\newcommand{\aap}{A\&A}

\newcommand{\placetabone}{
\begin{table}
\caption{Global parameters of the observed galaxies. Tabulated are the
  NGC number, the morphological type from NED, the nuclear activity class
  from Ho, Filippenko \& Sargent (1997), and the distance $D$ and the 
  physical scales both from Knapen et al. (2006). }
\label{tab:GlobPars}
\begin{center}
\begin{tabular}{llllll} 

\hline
   Galaxy     & Type        & Activity  & $D$   & Scale      \\
              &	            &           & (Mpc) & (pc/arcsec)\\
\hline
 NGC~\,\,473  & SAB(r)0/a   &           & 29.8  & 144.5      \\
 NGC~4314     & SB(rs)a     & L2        &  9.7  & 47.0    \\
 NGC~4321     & SAB(s)bc    & T2        & 16.8  & 81.4    \\
 NGC~5248     & (R)SB(rs)bc & H         & 22.7  & 110.1   \\
 NGC~5383     & (R')SB(rs)b & H         & 37.8  & 183.3   \\
 NGC~6951     & SAB(rs)bc   & S2        & 24.1  & 116.8   \\
 NGC~7217     & (R)SA(r)ab  & L2        & 16.0  & 77.6    \\
 NGC~7742     & SA(r)b      & T2/L2     & 22.2  & 107.6   \\
 \hline
\end{tabular}
\end{center}
\end{table}
}

\newcommand{\placetabtwo}{
\begin{table}
\caption{Observing run details. Given are the galaxy name, the date of
  the observations, the exposure times, and the position angles of the
  spectrograph slit, measured N over E.} 
\label{tab:ObsDetails}
\begin{center}
\begin{tabular}{llll} 

\hline
   Galaxy & Date & Exp. Times & PA\\
             &    & (s) & ($^\circ$)\\
\hline
 NGC~473  & 05/10/05 & 3$\times$1200 & 0\\
          & 05/10/05 & 3$\times$1200 & 65\\	
 NGC~4314 & 29/04/01 & 2$\times$1800 & 11\\
          & 29/04/01 & 3$\times$1800 & 90\\
 NGC~4321 & 30/04/01 & 3$\times$1200 & 127\\
          & 30/04/01 & 2$\times$1200 & 210\\
 NGC~5248 & 29/04/01 & 2$\times$1800 & 22\\
          & 29/04/01 & 1$\times$1800 & 136\\
NGC~5383 & 30/04/01 &  3$\times$1200 & 125\\
NGC~6951 & 29/04/01 & 2$\times$1800 & 150\\
		  & 05/10/05 & 3$\times$1200 & 19\\
		  & 05/10/05 & 3$\times$1200 & 110\\
 NGC~7217 & 05/10/05 & 3$\times$1200 & 0\\
 NGC~7742 & 05/10/05 & 3$\times$1200 & 22\\
          & 05/10/05 & 3$\times$1200 & 70\\
          & 05/10/05 & 3$\times$1200 & 115\\
\hline
\end{tabular}
\end{center}
\end{table}
}

\newcommand{\placefigone}{
\begin{figure*}
\begin{center}
\includegraphics[width=0.95\textwidth]{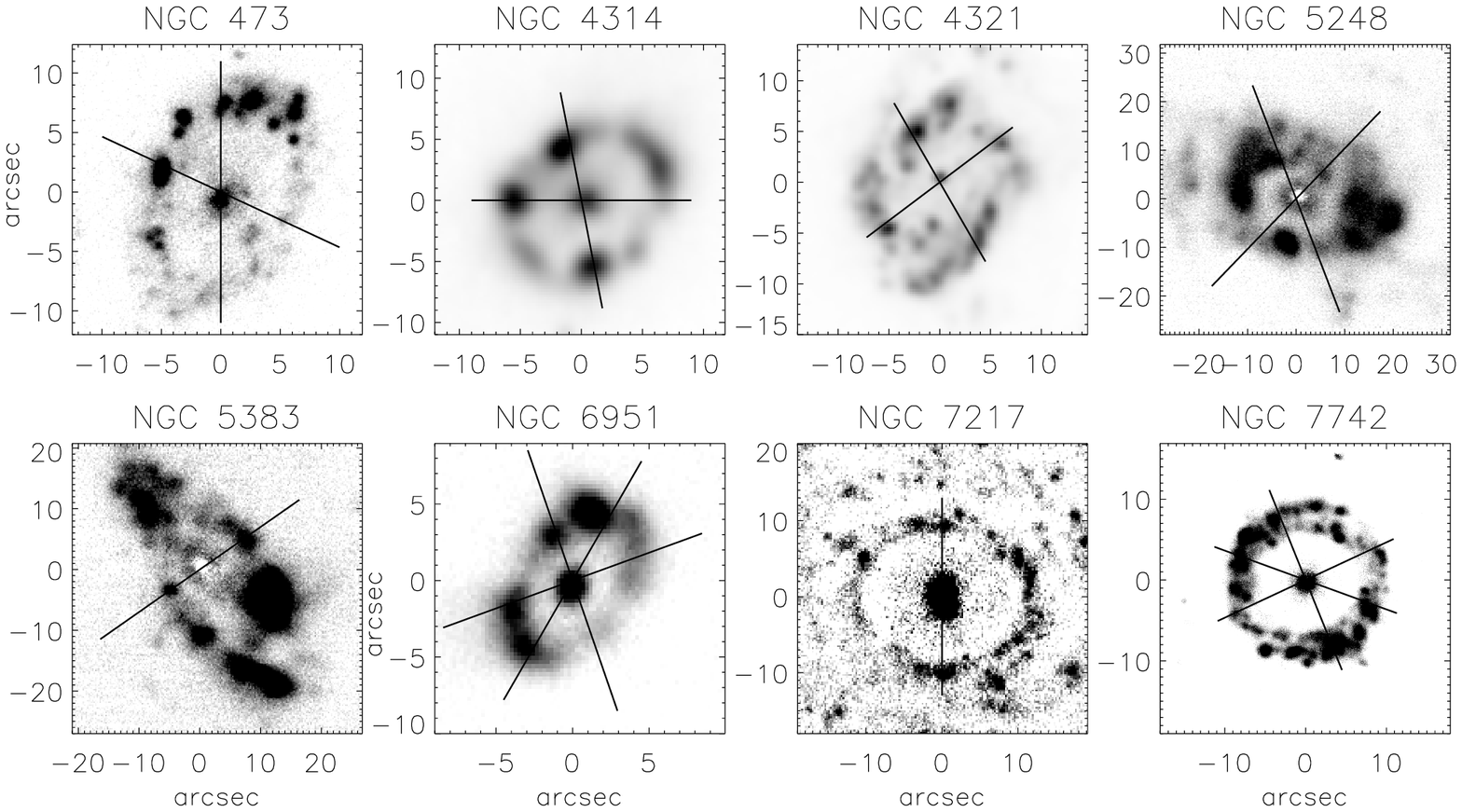}
\caption{
H$\alpha$ continuum-subtracted images of the nuclear rings in the
sample (data from Knapen et al. 2006). The black lines show the slit
positions of the spectrograph. North is up, East to the left.}
\label{fig:HalphaImages}
\end{center}
\end{figure*}
}

\newcommand{\placefigtwo}{
\begin{figure}
\begin{center}
\includegraphics[width=0.95\columnwidth]{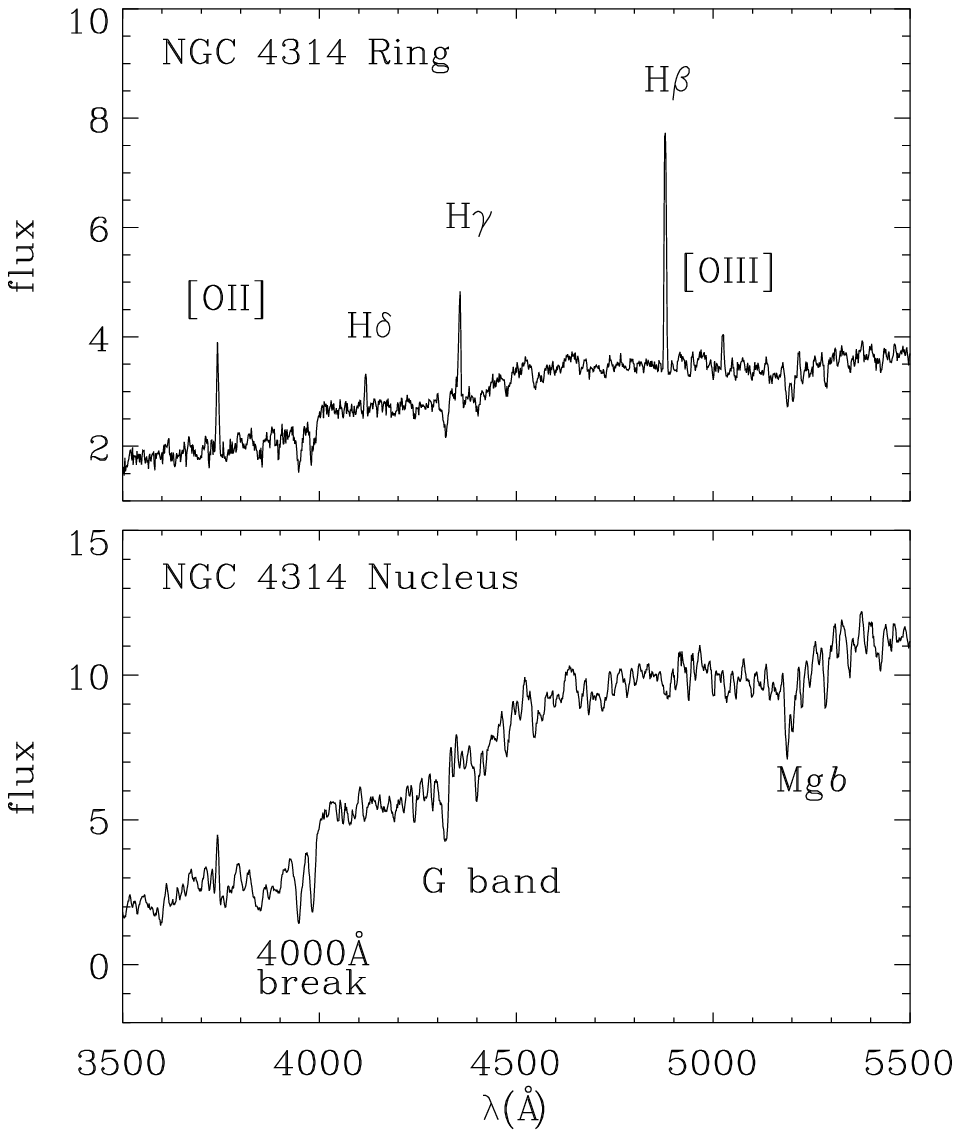}
\caption{
Two examples of the blue spectra obtained, one from the nuclear ring
($top$) and the other from the nucleus ($bottom$) of NGC~4314. The
flux scale is in units of $10^{-17}$\,ergs\,cm$^{-2}$\,s$^{-1}$\,\AA
$^{-1}$. Some of the emission and absorption features used in the
analysis are identified.}
\label{fig:SpecExamples}
\end{center}
\end{figure}
}

\newcommand{\placefigthree}{
\begin{figure*}
\begin{center}
\centering
\includegraphics[width=0.95\textwidth]{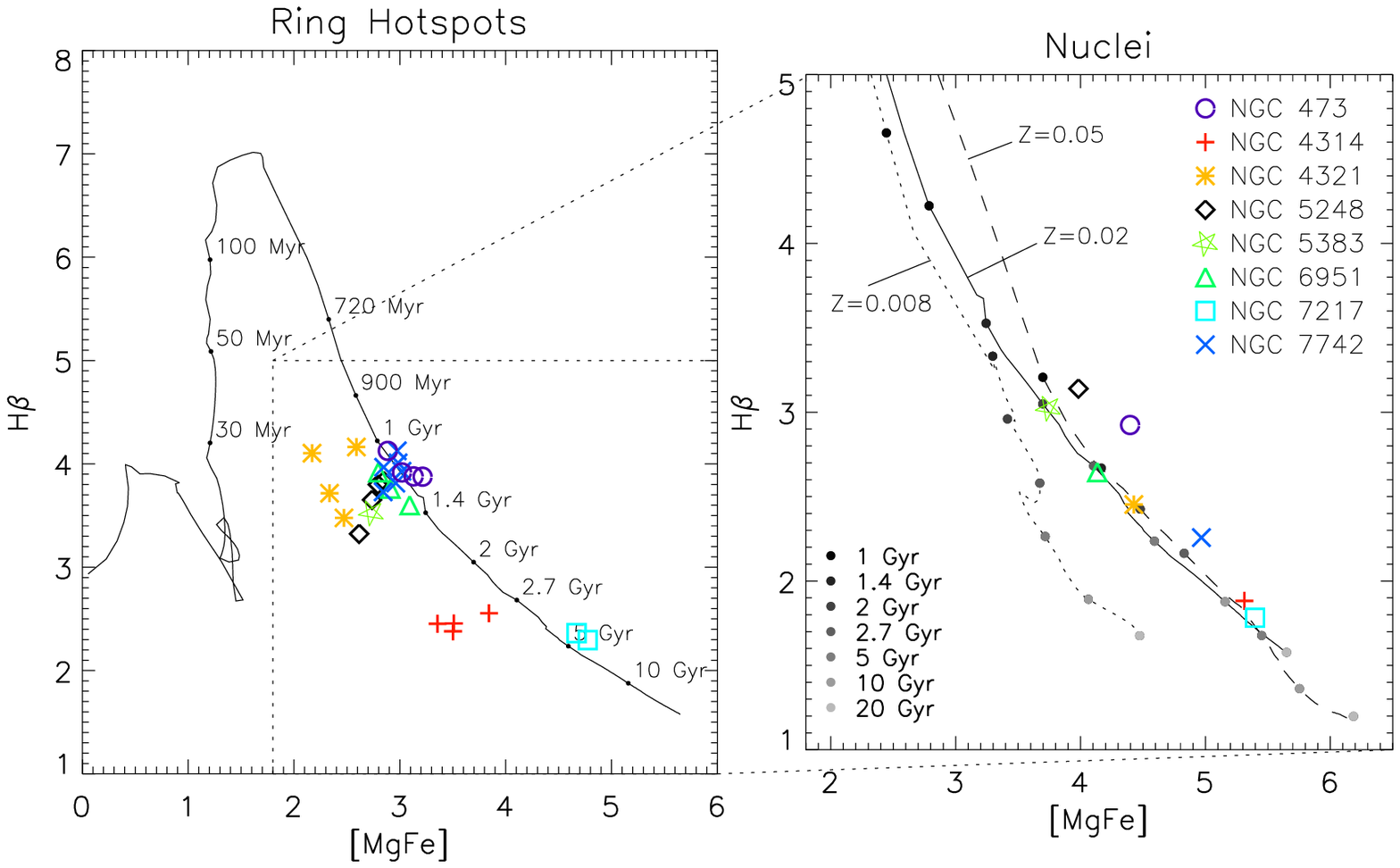}
\caption{
H$\beta$ vs. [MgFe] diagram for the ring hotspots and the nuclear
regions of our sample galaxies. {\it Left}: H$\beta$ and [MgFe] line
indices for the ring hotspots compared with the predicted time
evolution of the same indices for an single starburst stellar
population (SSP) of solar metallicity. {\it Right}: Same as left but
showing the region in the H$\beta$ vs. [MgFe] diagram occupied by our
nuclear measurements, together with SSP predictions for three
different metallicities, as labelled.}
\label{fig:IndexIndexDiagrams}
\end{center}
\end{figure*}
}

\newcommand{\placefigfour}{
\begin{figure*}
\begin{center}
\centering
\includegraphics[width=0.95\textwidth]{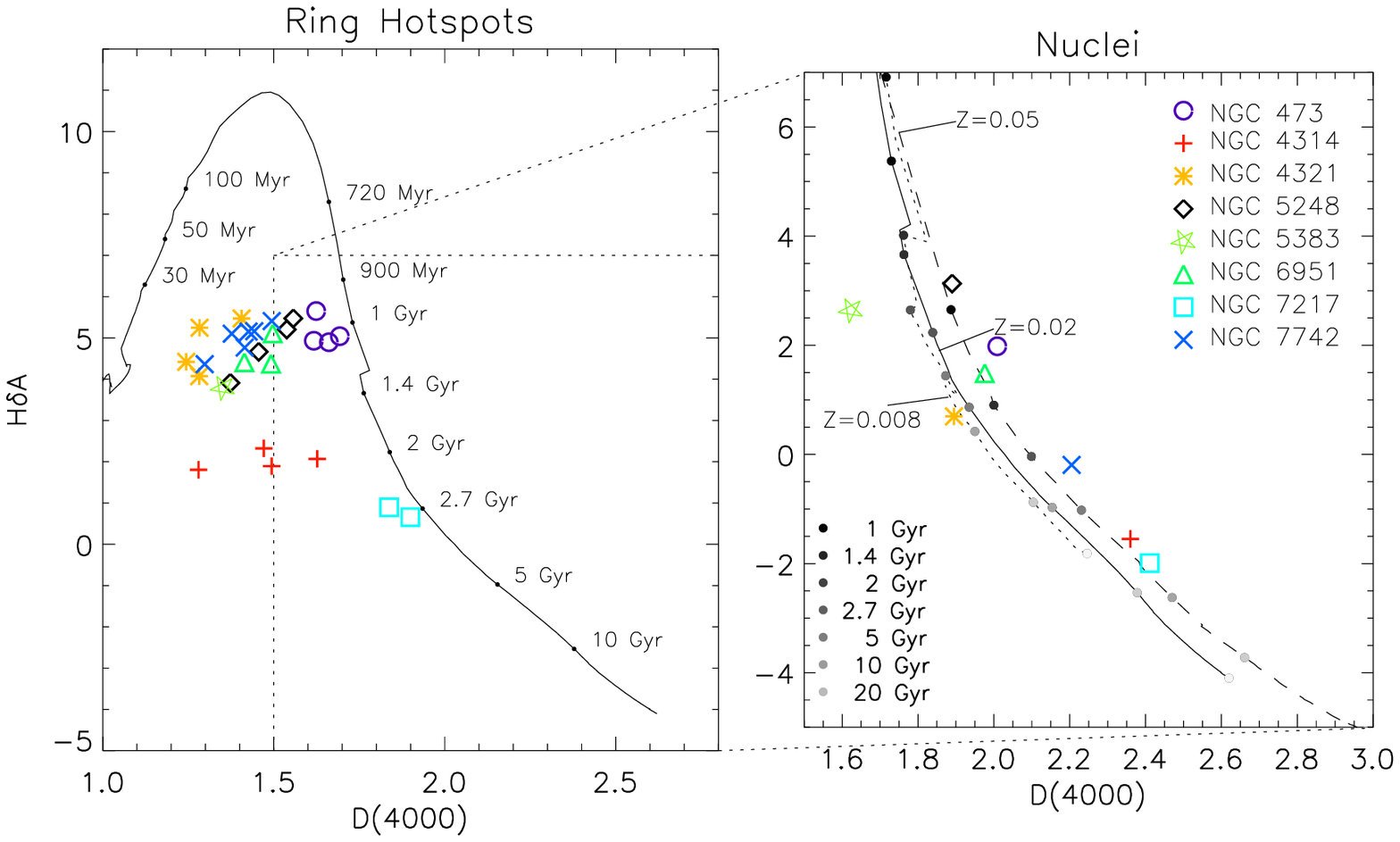}
\caption{
Same as Fig.~\ref{fig:IndexIndexDiagrams}, but now showing the \HdA\
and D(4000) indices. Notice the more pronounced discrepancy between
the position of the ring hotspots measurements and the predictions of
the single starburst model.}
\label{fig:IndexIndexDiagrams2}
\end{center}
\end{figure*}
}

\newcommand{\placefigfive}{
\begin{figure*}
\begin{center}
\centering
\includegraphics[width=\textwidth,bbllx=10pt,bblly=2pt,bburx=504pt,bbury=334pt]{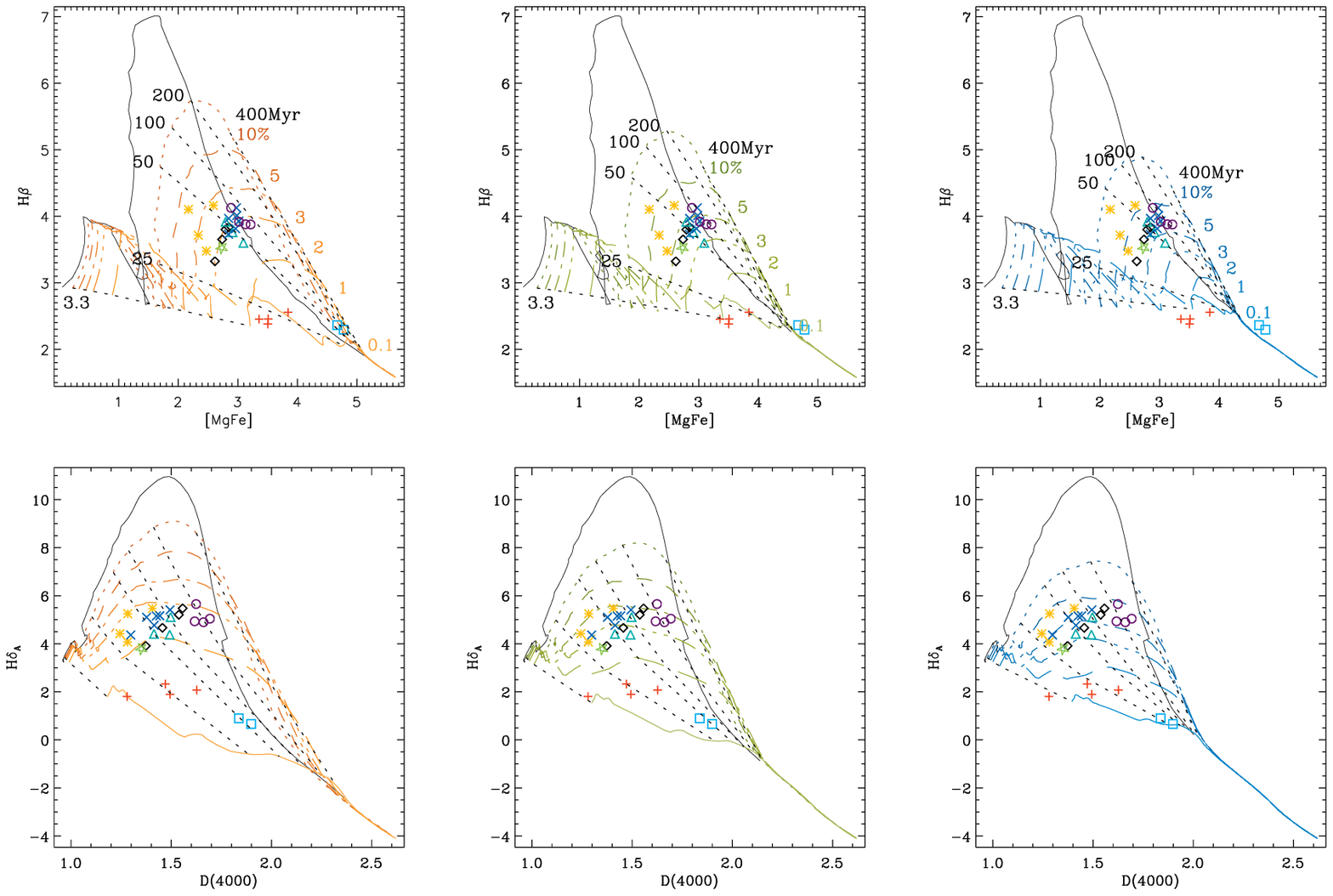}
\caption{
Composite stellar population models for the ring hotspots, assuming
the superposition of an old intervening/underlying bulge/disc
population and of young ring stars, the latter having formed in a
single star-formation episode. Both old and young components are
assumed to form instantaneously. The upper panels show the evolution
of the [MgFe] line-strength index vs. that of \Hb, whereas the lower
panels show the evolution of D(4000) vs. \HdA. From left to right the
models include progressively younger disc and bulge populations, which
formed 10, 5, and 3\,Gyr before the stars in the ring.  In all panels
the solid black curve shows, similarly to
Figures~\ref{fig:IndexIndexDiagrams} and
\ref{fig:IndexIndexDiagrams2}, the time evolution of the absorption
features in the old bulge and disc population, whereas the coloured
curves show the evolution of the same features in the hotspot spectra
after the onset of star formation in the ring.  In each panel,
differently coloured lines show how the strength of the ring starburst
is varied, adding from 0.1\% to 10\% of the mass in the older bulge
and disc population, as indentified in the top panels.  The position
of the hotspots data points are shown by the coloured symbols, as in
Figures~\ref{fig:IndexIndexDiagrams} and
\ref{fig:IndexIndexDiagrams2}.}
\label{fig:SingleBurst}
\end{center}
\end{figure*}
}

\newcommand{\placefigsix}{
\begin{figure*}
\begin{center}
\centering
\includegraphics[width=\textwidth,bbllx=10pt,bblly=2pt,bburx=504pt,bbury=334pt]{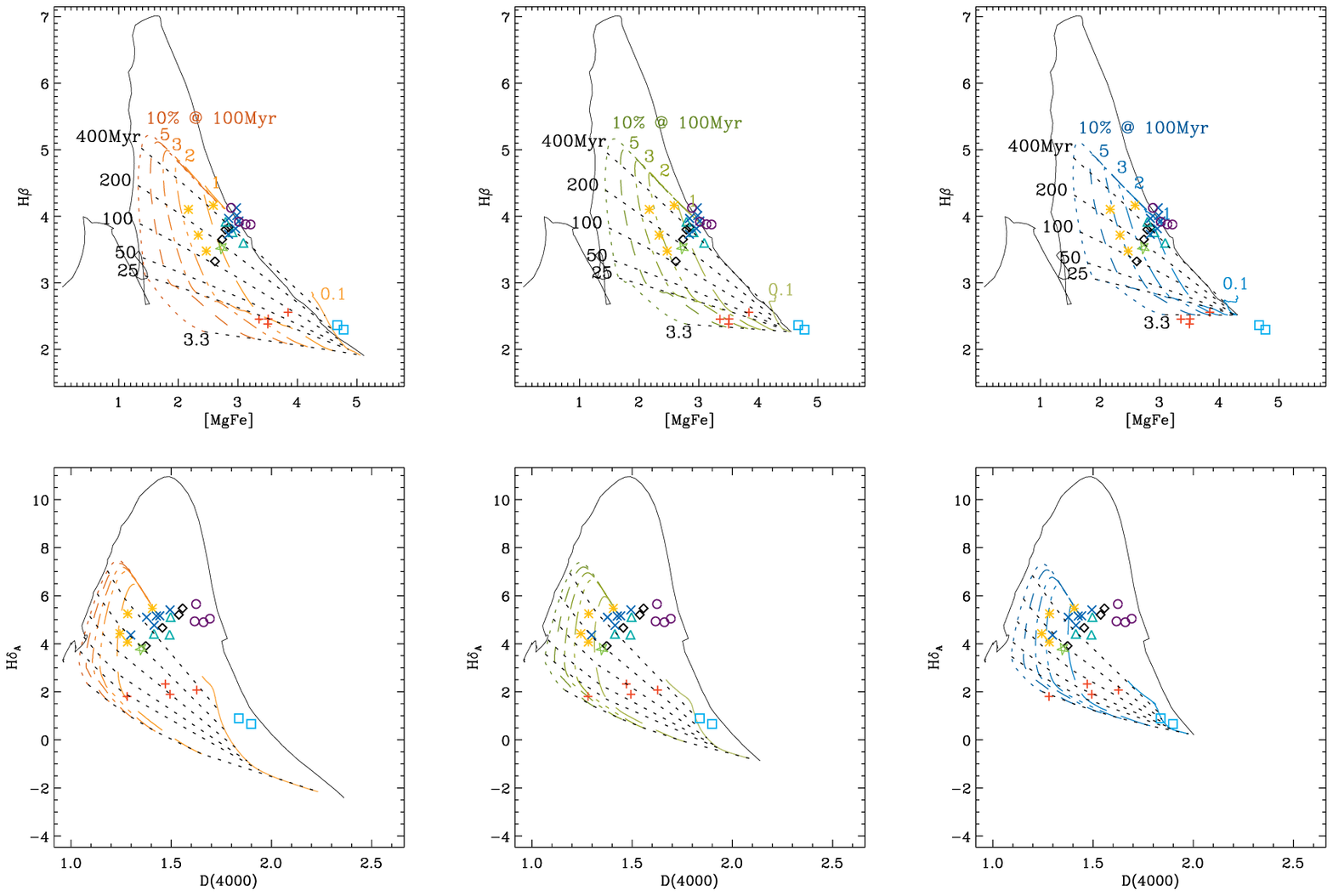}
\caption{
As Figure~\ref{fig:SingleBurst}, but now assuming that the ring stars
formed over a period of constant star formation. The strength of this
activity is quantified by stellar mass that is built in the ring after
100 Myr, relative to the bulge and disc mass.}
\label{fig:ContinousBurst}
\end{center}
\end{figure*}
}

\newcommand{\placefigseven}{
\begin{figure*}
\begin{center}
\centering
\includegraphics[width=\textwidth,bbllx=10pt,bblly=2pt,bburx=504pt,bbury=334pt]{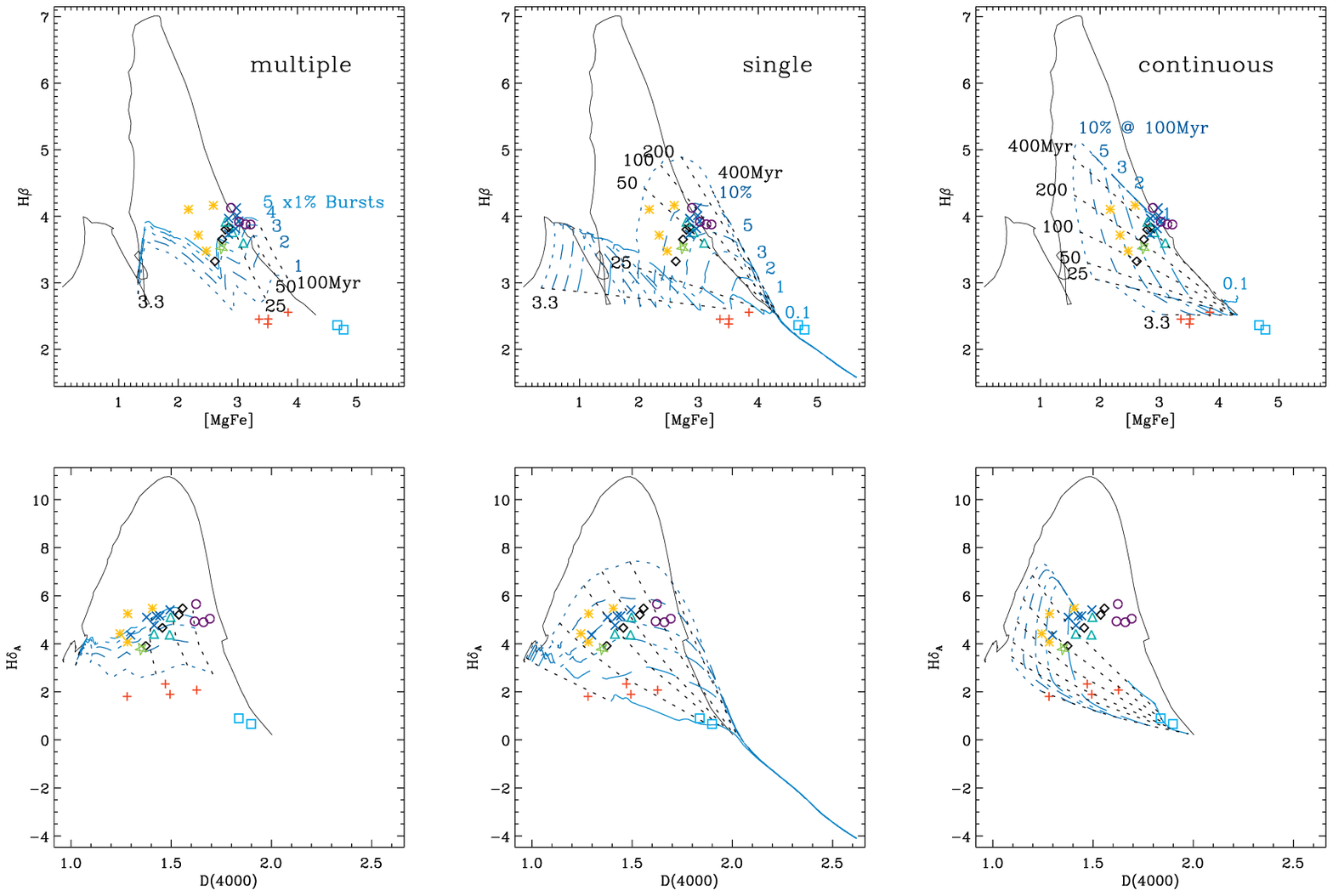}
\caption{
Composite stellar population models for the ring hotspots for the
superposition of a 3-Gyr-old bulge/disc population and of young ring
stars, which formed during a varying number of instantaneous
star-formation episodes (from 1 to 5 bursts, {\it left\/}), during a
single burst ({\it centre}), or over a period of constant star
formation ({\it right\/}). As in Figures~\ref{fig:SingleBurst} and
\ref{fig:ContinousBurst}, the top panels show the evolution of [MgFe]
vs. \Hb, whereas the lower panels show D(4000) vs. \HdA. The central
and right panels are similar to the right panels of Figures~5 and 6,
respectively. The multi-burst model ({\it left\/}) shows the time
evolution of the indices in the hotspot spectra since the onset of the
last starburst, each of which is set to contribute 1\% of the bulge
and disc mass.}
\label{fig:MultiBurst}
\end{center}
\end{figure*}
}

\newcommand{\placefigeight}{
\begin{figure}
\begin{center}
\includegraphics[width=0.95\columnwidth]{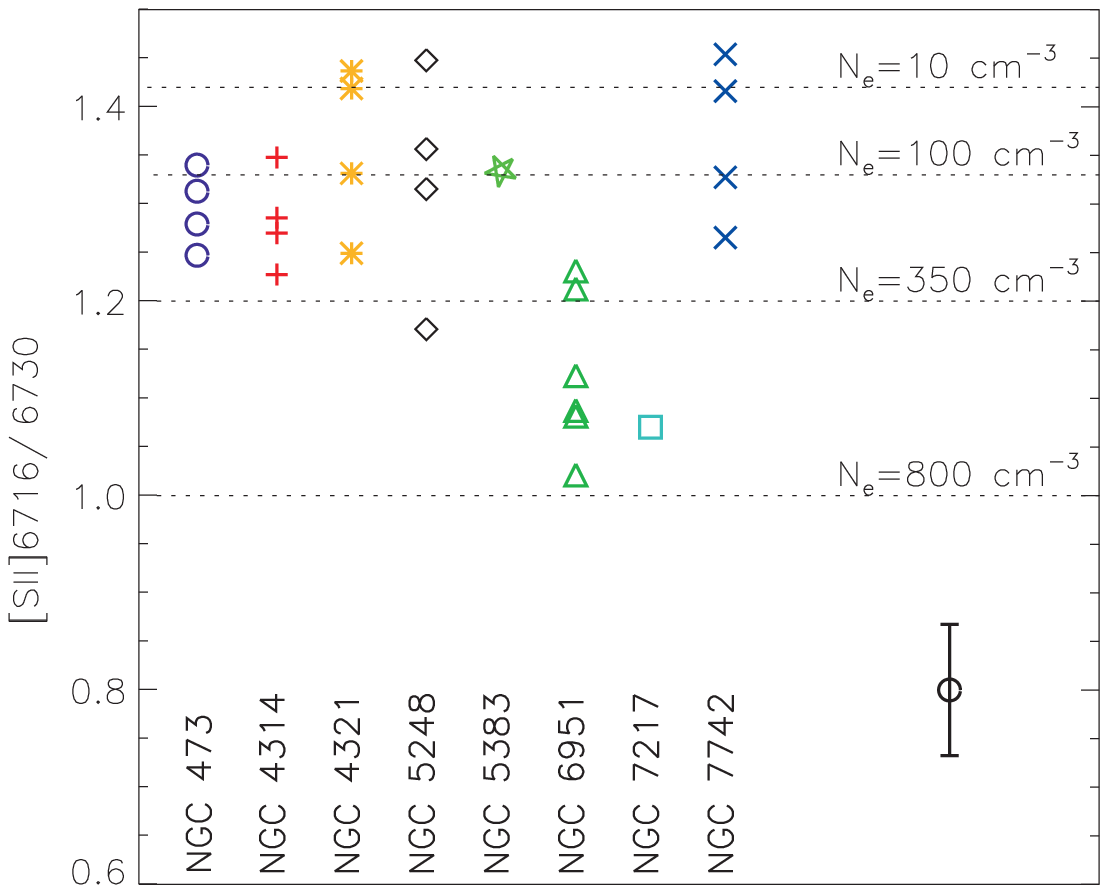}
\caption{
The \Sii$\lambda6716$/\Sii$\lambda6731$ ratio in the nuclear rings of
our sample galaxies. Assuming a temperature of 10$^4$\,K, the dashed
lines give estimates of the electron density. For NGC~7217, the two
hotspot spectra were combined to achieve a minimum signal to noise
ratio of 3 for the \Sii$\lambda6716$/\Sii$\lambda6731$ ratio. The
error bar shown at the bottom right of the plot is the average
uncertainty measured within our sample.}
\label{fig:SiiDensity}
\end{center}
\end{figure}
}

\newcommand{\placefignine}{
\begin{figure*}
\begin{center}
\includegraphics[width=\textwidth]{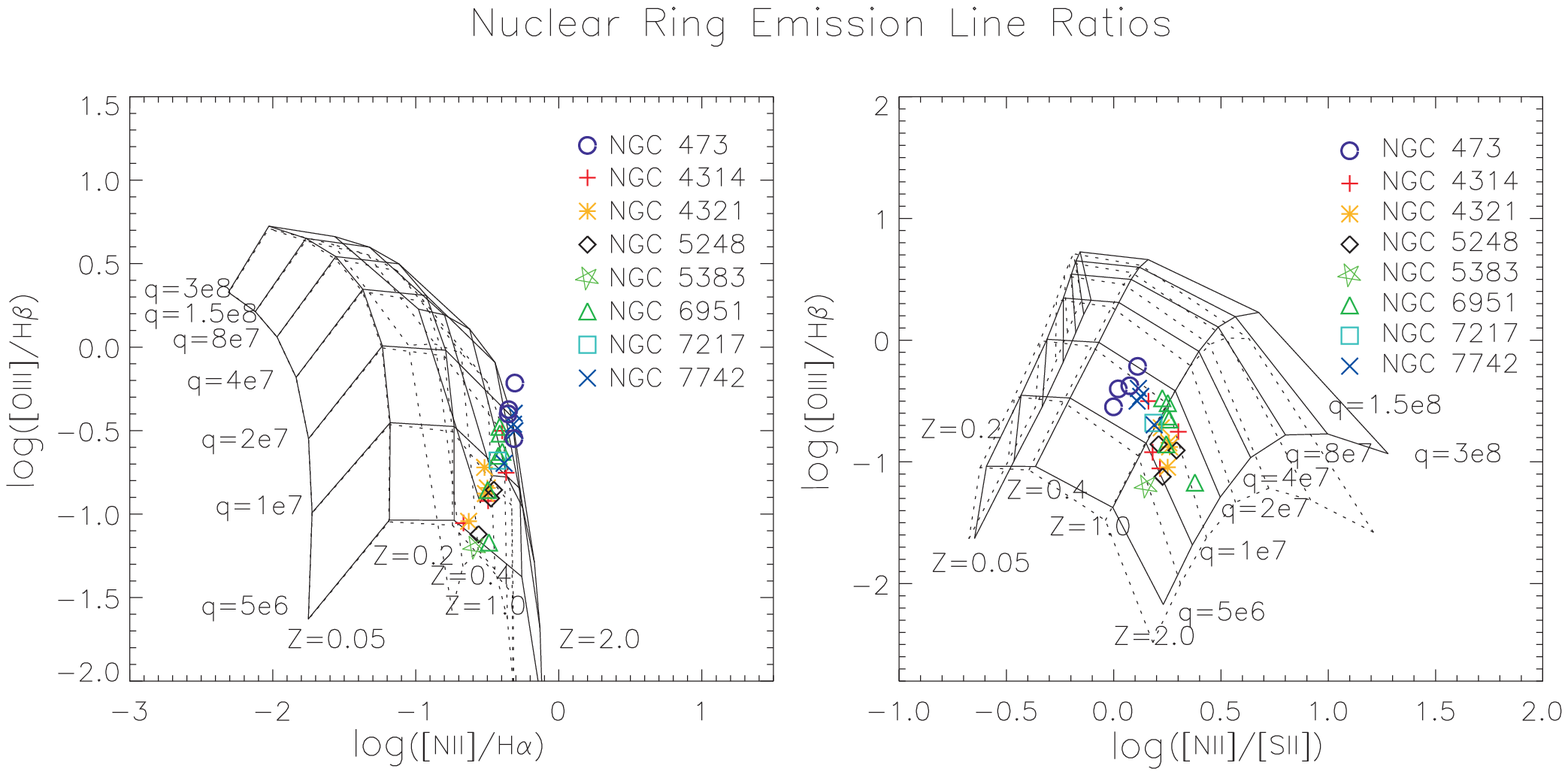}
\caption{
Diagnostic diagrams of log(\Oiii/\Hb) vs. log([\Nii/\Ha) ({\it
left\/}) and log(\Oiii/\Hb) vs. log(\Nii/\Sii) ({\it right\/}),
featuring the MAPPINGS III model prediction for the nebular emission
arising in star forming regions.  The models are based on STARBURST99
spectral energy distributions for continuous star formation over a
period of 8\,Myr. The theoretical grids assume different values for
the ionisation parameter and for the starburst metallicity, and are
shown with black lines for a density of 10\,cm$^{-3}$ and dashed lines
for a density of 350\,cm$^{-3}$. The data points measured from the
nuclear ring hotspots are shown as coloured symbols. Uncertainties in
the ratios are at most 30\%, translating into errorbars of 0.1 or less
along both the axes of these logarithmic plots.}
\label{fig:EmiLinesDiagnostic}
\end{center}
\end{figure*}
}

\newcommand{\placefigten}{
\begin{figure}
\begin{center}
\includegraphics[width=0.95\columnwidth]{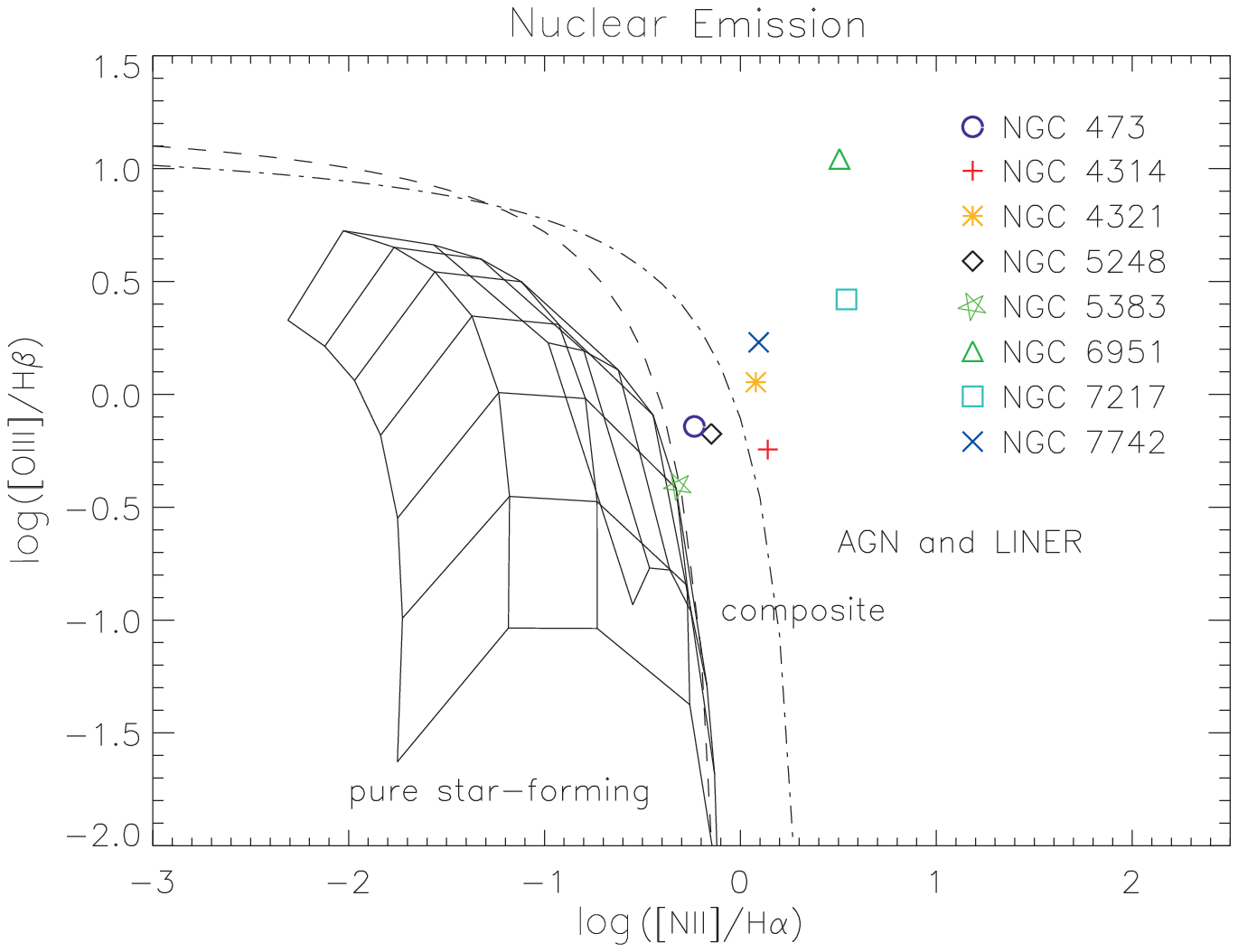}
\includegraphics[width=0.95\columnwidth]{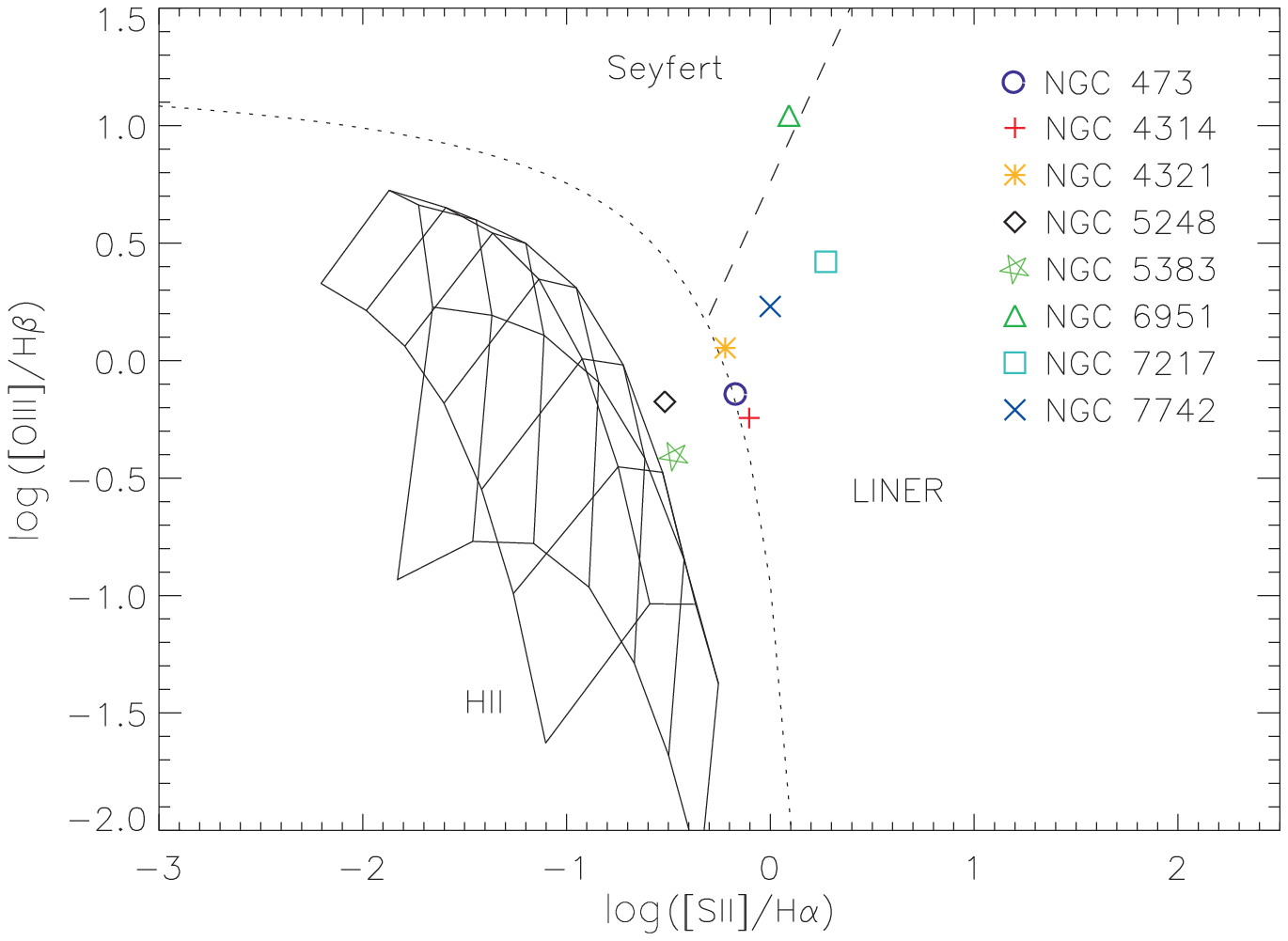}
\caption{
Diagnostic diagrams of log(\Oiii/H$\beta$) vs. log(\Nii/H$\alpha$)
({\it up\/}) and vs. log(\Sii/H$\alpha$) ({\it down\/}), compared with
theoretical model grids as in Fig.~\ref{fig:EmiLinesDiagnostic}. The
data points measured from the nucleus of each galaxy are plotted as
coloured symbols. {\it Up\/}: the dot-dashed line is the ``maximum
starburst line'' of Kewley et al. (2001). The dashed line is the
empirically derived starburst line from Kauffmann et al. (2003). {\it
Down\/}: the dotted line is the ``maximum starburst line'' of Kewley
et al. (2001), the dashed line from Kewley et al. (2006), to separate
Seyferts and LINERs.}
\label{fig:EmiLinesDiagnosticNucleus}
\end{center}
\end{figure}
}

\title[Star formation and stellar populations across nuclear rings in
galaxies] {Star formation and stellar populations across nuclear rings
in galaxies} 

\author[M. Sarzi et al.]{M. Sarzi$^{1}$\thanks{E-mail:
m.sarzi@herts.ac.uk}, E. L. Allard$^{1}$, J. H. Knapen$^{1,2}$ and
L. M. Mazzuca$^{3}$
\\ $^{1}$Centre for Astrophysics Research, University of
Hertfordshire, College Lane, Hatfield, Herts, AL10 9AB, UK
\\ $^{2}$ Instituto de Astrof\'\i sica de Canarias, E-38200 La Laguna,
Spain
\\ $^{3}$NASA Goddard Space Flight Center, Code 441, Greenbelt, MD
20771, USA}

\begin{document}

\date{Accepted 2007 July 2. Received 2007 June 22; in original form 2007 May 11}

\pagerange{\pageref{firstpage}--\pageref{lastpage}} \pubyear{2007}

\maketitle

\label{firstpage}

\begin{abstract}

We present a study of the optical spectra of a sample of eight
star-forming nuclear rings and the nuclei of their host galaxies.
The spectra were obtained with the ISIS spectrograph on the William
Herschel Telescope and cover a wide range in wavelength, enabling the
measurement of several stellar absorption features and gas emission
lines.

We compared the strength of the absorption lines to a variety of
population synthesis models for the star-formation history in the
nuclear rings, including also the contribution of the older bulge and
disc stellar components.
In agreement with our previous investigation of the nuclear ring of
NGC~4321, which was based on a more restricted number of line-strength
indices, we find that the stars in our sample of nuclear rings have
most likely formed over a prolonged period of time characterised by
episodic bursts of star-formation activity. Constant star formation is
firmly ruled out by the present data, whereas a one-off formation
event is an unlikely explanation for a common galactic component such
as nuclear rings. The nuclear rings of NGC~4314 and NGC~7217 have
distinct line-strength properties that set them apart from the rest of
our sample, which are due to a larger contibution of bulge star in the
observed spectra and, in the case of NGC~4314 to a younger stellar
population in the ring.

We have used emission-line measurements to constrain the physical
conditions of the ionised gas within the rings, using the ratio of the
\Sii$\lambda\lambda6716,6731$ lines to estimate the density of the
gas, and photoionisation model grids within specific diagnostic
diagrams to derive metallicity. We find that emission in all nuclear
rings originates from \Hii\ regions with electron densities typical
for these kinds of objects, and that the rings are characterised by
values for the gas metallicity ranging from slightly below to just
above Solar.
We have also studied the spectra of the nuclei of our sample galaxies,
all of which display emission lines. Consistent with previous studies
for the nuclear activity and stellar populations, the majority of our
nuclei appear to be dominated by old stellar populations and by
LINER-like emission.

As 20\% of nearby spiral galaxies hosts nuclear rings that are
currently forming massive stars, our finding of an episodic star
formation history in nuclear rings implies that a significant
population remains to be identified of young nuclear rings that are
not currently in a massive star formation phase. Nuclear rings may
thus be a much more common galactic component than currently known.

\end{abstract}


\section{Introduction}

Star-forming nuclear rings occur in approximately 20\% of spiral
galaxies (Knapen 2005). Generally, they are sites of intense active
star formation, containing a mixture of gas, dust and young
stars. This is deduced from their blue colours, bright hydrogen
emission lines in optical and infrared, enhanced radio continuum
emission, and overall patchy appearance (Arsenault et al. 1988; Pogge
1989; Garc\'{\i}a-Barreto et al. 1991; Benedict et al. 1992, 1996;
Knapen et al. 1995a,b; Jogee et al. 2002; Allard Peletier, \& Knapen
2005; Mazzuca et al. 2006, 2007). Additionally, red nuclear dust rings
have been identified (e.g., Buta \& Crocker 1991; Wozniak et al. 1995;
Vila-Vilar\'o et al. 1995), as well as so-called `fossil rings', which
are dust-free and composed entirely of stars, and virtually
indistinguishable from the surrounding stellar population (Erwin et
al. 2001; Erwin \& Sparke 2002).

The vast majority of nuclear rings appear in barred galaxies (Knapen
2005), where the bar has the ability to channel gas from the disc
towards the nucleus (e.g., Huntley, Sanders \& Roberts 1978; Sanders
\& Tubbs 1980; Simkin, Su \& Schwarz 1980; van Albada \& Roberts 1981;
Schwarz 1984; Combes \& Gerin 1985; Shlosman, Begelman \& Frank 1990;
Athanassoula 1992; Knapen et al. 1995a,b). During this process, the
radial inflow of gas is slowed down near the location of the inner
Lindblad resonances (ILRs), which allows for the accumulation of gas
and which can, ultimately, lead to a star-forming ring (Combes \&
Gerin 1985; Shlosman et al. 1989; Wada \& Habe 1992; Heller \&
Shlosman 1994; Knapen et al. 1995a,b; Fathi et al. 2005).
In a few cases, two of which will be discussed in this paper, nuclear
rings occur in host galaxies without an obvious kpc-scale bar. The
formation of such rings may be related to the presence of a weak oval
distortion (Buta et al. 1995), to a dissolved bar (Verdes-Montenegro
et al. 1995), to the tidal effects of a companion galaxy (Combes
1998), or to a recent minor merger (Knapen et al. 2004; Mazzuca et
al. 2006).

At least two thirds of all spirals are barred (de Vaucouleurs 1963;
Sellwood \& Wilkinson 1993; Knapen et al. 2000, Eskridge et al. 2000,
Gr$\o$sbol et al. 2004). Jogee et al. (2004), Elmegreen, Elmegreen \&
Hirst (2004) and Zheng et al. (2005) showed that the bar fraction is
roughly constant out to $z \approx 1$, indicating that bars are
long-lived phenomena. Since gas is readily available in the disc,
nuclear rings may continue to form stars as long as the bar continues
to provide an inflow of fresh material. Even with continued inflow at
the rates thought to be supplied by a typical bar (around a Solar mass
per year), the high star formation rates within the rings can exhaust
the gas relatively quickly (Elmegreen et al. 1998), effectively
turning off the star formation. As more gas is supplied to the ring
the density will gradually increase, until star formation is initiated
again. The moderate inflow rates and observed elevated star formation
rates thus indicate that the star-forming process in nuclear rings may
cycle through active and quiescent periods. The fraction of galaxies
with nuclear rings may thus be significantly larger than the fraction
of rings seen to be forming massive stars at present (which is around
20\%; Knapen 2005).

At the current time, however, observational confirmation of this
longevity of nuclear rings is still limited. As such, the effect
nuclear rings may have on the evolution of their host galaxies is
unclear. 
The ages of individual star-forming regions, sometimes referred to as
hot spots, in nuclear rings have been studied using a variety of
techniques, almost all based on the use of optical or near-IR
recombination emission lines via imaging or spectroscopy (e.g.,
Engelbracht et al. 1998; Ryder \& Knapen 1999; Puxley \& Brand 1999;
Kotilainen et al. 2000; Alonso-Herrero, Ryder \& Knapen 2001; Ryder,
Knapen \& Takamiya 2001; D{\'{\i}}az-Santos et al. 2007; Mazzuca et
al. 2007). These lines trace primarily very young and massive stars,
and as a consequence the ages derived using these techniques are
usually below 10\,Myr, or otherwise very sensitive to assumptions on
the star-formation history.
Broad-band colours have also been used to estimate the age of the
stellar population in circumnuclear rings (e.g., Kotilainen et
al. 2000; Harris et al. 2001). Such measurements, however, are
affected by reddening and degeneracies between age and metallicities,
and even in the case of near-IR data, where the impact of dust is
mitigated, the interpretation of colour-colour diagrams is complicated
by the fact that populations of very different ages have very similar
colours (see, e.g., the models of Maraston 2005).
In contrast, the technique employed in the present paper, which is
based on the strength of several stellar absorption lines, is
sensitive to a much larger range of intermediate ages, is unaffected
by reddening and is known to break the age-metallicity degeneracy that
limits broad-band estimations (Worthey 1994).

Based on such an interpretation of line-strength indices, in Allard et
al. (2006, hereafter Paper I) we presented strong evidence that the
nuclear ring in NGC~4321 (Messier~100) is indeed long-lived, though
forming stars episodically in a rapid succession of bursts rather than
continuously.  Due to the limited wavelength range of the data we
used, we could, however, not completely rule out the possibility that
the rings are transient objects. With our long-slit data (presented in
this paper), we cover a much wider wavelength range and aim to confirm
the nature of the nuclear ring in NGC~4321, as well as seven other
well known rings, with the goal of generalising the star formation
activity pattern and evolutionary past within nuclear rings.

In the present paper we study the star formation histories of eight
nuclear rings, by extracting information from longslit ISIS
(Intermediate dispersion Spectrograph and Imaging System) spectra over
a wide range in wavelength, analysing both the stellar and gaseous
contributions to the spectra. For each nuclear ring, we have one to
three spectrograph slit position angles (PAs) which each bisect the
nucleus and the ring, yielding data on two to six \Hii\ regions,
referred to as hotspots, for each ring (see
Figure~\ref{fig:HalphaImages}).  Studying the stellar absorption and
gas emission lines will thus allow us to determine the physical
parameters and complex star formation properties of these rings.

The paper is structured as follows. Section~\ref{sec:ObsDataRedu}
details the observations and data reduction, and
Section~\ref{sec:Analysis} describes the analysis of the data. The
results of the stellar absorption line analysis are presented in
Section~\ref{sec:AbsLineDiagn}, and the analysis of the emission lines
is presented in Section~\ref{sec:EmiLineDiagn}. Finally, we present
our discussions and conclusions in Sections~\ref{sec:Discussion} and
\ref{sec:Conclusions}.

\placetabone
\placetabtwo
\placefigone
\placefigtwo

\section{Observations and Data Reduction}
\label{sec:ObsDataRedu}

The observed galaxies were chosen from the \Ha\ sample of Knapen et
al. (2006), and are all known to possess strong H$\alpha$-emitting
nuclear rings. Our sample consists of six galaxies with moderate to
strong bars, and two galaxies that appear to be unbarred, at least at
first glance. These two unbarred galaxies, NGC~7217 and NGC~7742, are
both known to have counter-rotating sub-systems in their central
regions (Merrifield \& Kuijken 1994; Sil'chenko \& Afanasiev 2000; de
Zeeuw et al. 2002), which may suggest a history of interaction. Some
global parameters for the eight galaxies are listed in
Table~\ref{tab:GlobPars}. The ISIS spectrograph on the William
Herschel Telescope (WHT) on La Palma was used to observe five galaxies
on 2001 April 29 and 30. An additional three galaxies were observed on
the 2005 October 5 using the same instrument. Details of the
observations, including spectrograph slit PAs for each galaxy, are
listed in Table~\ref{tab:ObsDetails}. \Ha\ images of all the nuclear
rings and the slit PAs used in each case are shown in
Figure~\ref{fig:HalphaImages}.

The slit was always positioned so it targetted the nucleus and at
least one bright hotspot in the ring. A slit width of 1 arcsec was
used. We operated ISIS with a dichroic, which allows simultaneous
observations with a red and a blue arm, covering wavelength ranges of
3500-5700\,\AA\, and 5900-8900\,\AA, respectively. The R158B grating
was used for the blue arm, providing 1.62\,\AA/pixel and
0.2\,arcsec/pixel scales with the EEV12 detector. For the initial
observing run the R158R grating was used with the TEK4 detector on the
red arm, which gave 2.90\,\AA/pixel and 0.36\,arcsec/pixel scales. For
the second observing run the TEK4 detector was no longer in operation,
and the red MARCONI detector was used, which provides similar spatial
and spectral scales to the blue EEV12 detector. The instrumental
resolution was 180 \kms\ ($\sigma=6.8$\,\AA) for the blue, and 115
\kms\ ($\sigma=5.8$\,\AA) for the red arm. Standard CuAr+CuNe
calibration lamp exposures were taken before and after each science
exposure, along with a number of spectroscopic standard star
observations for flux calibration.

Although the spectrograph slit passed through two hotspots in
NGC~5383, neither had high enough signal to noise ratio to be
useful. The two extracted spectra were instead combined. Additionally,
the two hotspots measured in NGC~7217 had sufficient signal to noise
in their stellar continuum, but not in their gas emission lines. The
two spectra were thus averaged for the emission line measurements but
not for the stellar index measurements.

The spectra were reduced within the {\sc IRAF} package using standard
methods. The data were first bias and overscan corrected, before
trimming and flatfielding. A cosmic ray removal algorithm was applied
next. The separate exposures were then checked for offsets in
positions between adjacent exposures, shifted if necessary, and
combined. The spatial regions along the slit containing the hotspots
and the nucleus were extracted and median-combined to produce a
one-dimensional spectrum. CuNe+CuAr lamp exposures were used for
wavelength calibration, with typical errors of 0.1\,\AA. The spectra
were traced to align the dispersion and spatial axes with the rows and
columns of the CCD array. To remove the background sky contribution,
regions at the outer edges of the slit were averaged to produce a
one-dimensional spectrum which was subtracted from every row in the
image. The spectra were flux calibrated to an absolute scale with
units of ergs\,cm$^{-2}$\,s$^{-1}$\,\AA$^{-1}$, with typical errors of
30\%.  A number of atmospheric absorption and emission lines is found
towards the end of the red spectral range, which could not be removed
sufficiently in all cases. Consequently, the red spectra were trimmed
to give a wavelength coverage of 5800-6800\AA, which includes the
important emission lines of H$\alpha$, \Nii\ and \Sii.

Figure~\ref{fig:SpecExamples} shows two examples of blue spectra
subtending the nucleus and one of the ring hotspots of NGC~4314.

\section{Data Analysis}
\label{sec:Analysis}

The stellar and gaseous contributions to the spectra were separated
using the direct fitting method described in Sarzi et
al. (2006)\footnote{We have used a modified version of the {\it
Gandalf\/} IDL code available at
{http://www.strw.leidenuniv.nl/sauron/}, to include also reddening by
dust.}.
We modelled the stellar continuum with linear combinations of
synthetic spectral templates from Bruzual \& Carlot (2003), adopting
the same library used by Tremonti et al. (2004). This includes 13
template spectra for each of the three metallicities $Z=0.004,0.02,
0.05$, corresponding to 10 instantaneous-burst models with ages from
5\,Myr to 11\,Gyr, a constant star formation model with an age of
6\,Gyr, and two models with exponentially declining star formation
histories with ages of 9 and 12\,Gyr.

The spectral regions affected by nebular emission are not excluded
from the fitting process. Instead, the emission lines are treated as
Gaussian templates and fitted simultaneously with the stellar
templates to the observed spectra. This has the advantage of
maximising the spectral information available to the fitting
algorithm.

As the blue and red spectra have different spectral resolutions they
could not be fitted simultaneously. The blue spectra cover a much
wider wavelength range than the red, and contain a large number of
stellar absorption features that help constrain the mix of stellar
templates. We therefore fit the blue spectra first, and use the
resulting optimal combination of templates to match the red spectra,
allowing only for a different velocity broadening. The fits to the
blue spectra include reddening due to foreground interstellar dust,
both in the Milky Way and in the sample galaxies, and due to dust in
the emission-line regions. The latter affects only the fluxes of the
emission line templates, and is constrained by the expected and
observed decrement of the Balmer lines.
Setting physically motivated limits on the intensity of the emission
from high-order Balmer lines is of particular importance to ensure
that the strength of the corresponding absorption features remains
unbiased, and will thus not affect our stellar population analysis.
A similar methodology was applied also by Shields et al. (2007).
The derived weights of the stellar spectra that define the optimal
template are used to reconstruct an unconvolved optimal template,
which is free from the effects of velocity broadening.
This unconvolved optimal template is used to measure the absorption
line strengths.

Owing mostly to the varied nature of the Tremonti et al. library, the
template-fitting procedure yield excellent results in matching the
spectra extracted from both the ring and nuclear apertures in all but
two cases. The two exceptions are the blue spectra of NGC~6951 at
PA=110$^{\circ}$ and 150$^{\circ}$. In the following, we will ignore
the line strength measurements obtained from these spectra.  Since the
nebular emission is sufficiently strong to allow robust estimates of
the emission-line fluxes, even where the fit to the stellar continuum
is poor, we will use the emission line results derived from these two
spectra.

\placefigthree
\placefigfour

\subsection{Line strengths and indices}
\label{subsec:Analysis_indices}

Stellar absorption lines provide information on the stellar content of
the rings. The strength of an absorption line index is determined by
the difference between the absorption line and the continuum level. An
index is typically defined by a wavelength range that contains the
feature and two wavelength ranges to either side that provide an
estimate of the red and blue continuum levels. The midpoints of the
continuum passbands are fitted by a linear relation, and the
difference between this and the absorption feature provides the index
value. The latter is sensitive to the shape of the continuum, the
spectral resolution, and the velocity dispersion of the object. All
these factors must be taken into account before one set of
measurements can be compared with another. The Lick/IDS system
(Worthey 1994) is commonly used to compare measurements from different
observations. For the data used in this paper, however, we are mostly
concerned with comparing our indices with model predictions, and for
this we need only ensure that our indices are measured in the same way
as they are in the models.
By construction, the model predictions come from model spectra with
identical resolution to our optimal templates, so it is appropriate to
compare them.
The line strengths were measured using {\sc indexf}, a C++ program
written by N. Cardiel\footnote{See
{http://www.ucm.es/info/Astrof/software/indexf/}}, with the Lick/IDS
wavelength definitions (but not calibrated to the Lick/IDS system).

\subsection{Uncertainties}
\label{subsec:Analysis_errors}

The uncertainties in the index measurements are largely determined by
the quality of the optimal template fit to the observed galaxy
continuum. Kuntschner et al. (2006) describe the errors associated
with our emission and absorption line fitting method and subsequent
measurements of line strengths. From a detailed analysis they find a
constant error budget of 0.1\,\AA\, from the continuum fitting
process, which we adopt for all the stellar absorption line indices
presented here.
With regard to the emission-line fluxes, Sarzi et al. (2006) find from
simulations that the uncertainties associated with flux measurements
scale with the level of fluctuations in the residuals of the fit to
the spectrum.
We convert such residual-noise levels into an estimate for the
emission-line flux errors, adopting as an error for each line the flux
of a Gaussian with amplitude equal to the residual-noise level and
width equal to the width of the emission-line in question.
Considering the weaker lines that will enter our emission-line
analysis, \Hb\ and \Oiii, these come on average with 3\% and 26\%
uncertainties, respectively, which translate to a typical error of
0.12 on the logarithm of the \Oiii/\Hb\ line ratio.
Errors on line ratios based on the stronger emission lines in our red
spectra (e.g., \Ha, \Nii, \Sii) will be smaller than that.

\placefigfive
\placefigsix 
\placefigseven

\section{Stellar Absorption Line Diagnostics: The Star Formation History}
\label{sec:AbsLineDiagn}

\subsection{Index-index diagrams and single starburst models}
\label{subsec:AbsLineDiagn_single}

To probe the properties of unresolved stellar populations, several
combinations of absorption-line strength indices are traditionally
used to separate age and metallicity effects, comparing the measured
values to predictions for stellar populations that were born in an
instantaneous burst of star formation and for a single metallicity.
Using such single starburst population (SSP) models from Bruzual \&
Charlot (2003), we plot the evolution of the \Hb\ and [MgFe]
($=\!\sqrt{\mbox{Mg{\emph b}}\times\mbox{Fe5015}}$; see
Falc\'on-Barroso 2002) indices with time for a solar metallicity
(Figure~\ref{fig:IndexIndexDiagrams}). The \Hb\ absorption index is
low in very young stars, increases with age until it peaks at around
100-200\,Myr, then decreases with age. Conversely, the [MgFe] index
remains almost constant for a young population while the \Hb\ index
increases, and increases monotonically with age after the strength of
\Hb\ has peaked.

The values of the \Hb\ and [MgFe] indices for the nuclear ring
hotspots in our sample galaxies are compared to the predicted time
evolution of these indices for SSP models in the left panel of
Figure~\ref{fig:IndexIndexDiagrams}.
While some of the data points lie close to the solar metallicity SSP
line, most do not, suggesting that a star formation history more
complex than an SSP must be responsible for the observed values. On
the other hand, the stellar populations in the nuclei of our sample
galaxies (Figure~\ref{fig:IndexIndexDiagrams}, right panel) are well
described by SSP models of old ages, and are consistent with previous
studies for the nuclear populations of bulges (Sarzi et al. 2005;
Gonz\'alez Delgado et al. 2004).
The star-forming nuclear rings are undoubtedly observed along the same
line-of-sight as an older stellar population in the disk and bulge of
their host galaxies. As a consequence, the scenario most likely to
match the line strengths of our hotspot spectra is that of a
superposition of young stars in the ring, and old stars in the
underlying disc and intervening bulge. This situation is similar to
the one described in Paper~I for NGC~4321, where we used the same
indices as shown in Figure~\ref{fig:IndexIndexDiagrams}.

The larger wavelength range of our current data set allows us to
analyse more indices than was possible in Paper~I, which enables us to
investigate the star formation history of the nuclear rings in more
detail.
In addition to \Hb\ and [MgFe], we have measured the \HdA\ (Worthey \&
Ottaviani 1997) and D(4000) (Bruzual 1983) indices in all the hotspot
and nuclear spectra. Like other Balmer indices, the \HdA\ index is an
age-sensitive index that is particularly strong in stellar populations
with ages 0.1-1\,Gyr. The advantage of using higher order Balmer
indices such as \HdA\, however, is that they are much less affected by
the presence of emission. The D(4000) index is derived from the ratio
of two passbands that lie on either side of the 4000\,\AA~break (see
Figure~2), which occurs as a large number of stellar absorption
features appear bluewards of this break in cooler, more opaque
stars. In hotter, younger stars, the 4000\,\AA~break is therefore
smaller than it is in cooler, older stars.

Figure~\ref{fig:IndexIndexDiagrams2} shows how our \HdA\ and D(4000)
measurements compare with the SSP predictions. In the D(4000)
vs. \HdA\ diagram {\it none\/} of the ring hostspot data can be fitted
by the SSP model, confirming the need for a combination of young and
old stars to explain the spectra of the nuclear ring regions. On the
other hand, the nuclear data remain consistent with SSPs.
The similarities between the results inferred from the two pairs of
absorption line indices confirm the reliability of the emission-line
correction.
We note that although SSP models with subsolar metallicities could
match the position of some of the hotspots data points in either of
the [MgFe] vs \Hb\ or D(4000) vs. \HdA\ diagrams, they could never do
so in both diagrams with the same value of the
metallicity. Furthermore, no SSP model can match the D(4000) index
values below 1.5 observed in many hotspots.

Using the SSP models as a baseline to interpret the index values for
the hotspots (as plotted in Figures~\ref{fig:IndexIndexDiagrams} and
\ref{fig:IndexIndexDiagrams2}) as a superposition of young and old
stellar populations, we can distinguish two separate groups of
objects.
The first group includes the majority of the hotspots, for which it
would seem plausible to match their position in
Figure~\ref{fig:IndexIndexDiagrams} and \ref{fig:IndexIndexDiagrams2}
by combining $\sim30-50$\,Myr young stars with disc and bulge stars
older than $\sim1$\,Gyr. On the other hand, considerably weaker \Hb\
and \HdA\ lines in the hotspots of NGC~4314 and NGC~7217 suggest a
mixture of even younger ring stars and a disc/bulge population older
than $\sim2.7$\,Gyr.
Assuming that the stellar populations within our nuclear apertures are
a good approximation for the underlying disc and bulge populations
encompassed by our hotspot spectra, the nuclear data points for
NGC~4314 and NGC~7217 further suggest the presence of a much older
underlying and intervening stellar population than in the case of our
remaining sample galaxies.

In the next section, we will compare the position of the hotspots in
the previous diagrams with detailed models featuring a combination of
a young ring population with an older disc and bulge stellar
population.
For simplicity, we will describe the disc and bulge stellar components
with a single SSP model.
We will allow for more complex star formation histories in the nuclear
rings of our sample galaxies than simple instantenous bursts,
considering also a continous star formation history and multiple
starburst episodes.

\subsection{Composite stellar population modelling}
\label{subsec:AbsLineDiagn_composite}

\subsubsection{Instantaneous star formation in the ring}
\label{subsubsec:AbsLineDiagn_instantaneous}

The simplest composite population model for the stellar populations at
the hotspot locations assumes that both the young ring population and
the older disc and bulge populations can be described by means of SSP
models.
Figure~\ref{fig:SingleBurst} shows several of such stellar population
models, following the evolution of the line strength indices ([MgFe]
vs. \Hb\ upper panels, D(4000) vs. \HdA\ lower panels) after the onset
of a single instantenous star formation episode in the ring. From left
to right the models include progressively younger disc and bulge
populations, which formed 10, 5, and 3\,Gyr before the stars in the
ring. In each panel, the strength of the ring starburst is varied,
adding from 0.1\% to 10\% of the mass in the older bulge and disc
population.

Starting from the first group of objects, the [MgFe] vs. \Hb\ diagrams
of Figure~\ref{fig:SingleBurst} indicate a rather coherent picture
where the majority of the hotspots data points appear to trace the
evolution of an SSP $\sim30-100$\,Myr after its formation. The figure
shows that the bursts contribute $\sim2$\%, 2-3\%, or 3-5\% of the
mass of the bulge and disc population, respectively, and indicate
slightly older ages for the higher mass fractions. The measurements
for NGC~4321 do not suggest such a clean evolution, however, although
they indicate a relatively more massive ring starburst.
Excluding this object, the D(4000) vs. \HdA\ diagrams yield a rather
similar picture for this group of hotspot spectra, in particular when
considering the models with a 3\,Gyr old disc and bulge population.
In this respect, we note that the template fitting for the hotspot
spectra of NGC~4321 was quite poor compared to the rest of the sample,
so that the line strength indices may be somewhat less reliable for
this galaxy.

For the hotspots in NGC~4314 and NGC~7217, a bulge and disc population
younger than 5\,Gyr is ruled out by the [MgFe] vs. \Hb\ diagrams of
Figure~\ref{fig:SingleBurst}. For a 10\,Gyr old disc and bulge
component, the ring starburst in these galaxies appears to have
occurred less than 50\,Myr ago and to have contributed less than 1\%
of the mass.
The case for a significantly younger ring population is particularly
strong in the case of NGC~4314, since the position of the NGC~7217
datapoints in the [MgFe] vs. \Hb\ diagram cannot deliver a robust age
estimate.
Considering that NGC~4314 and NGC~7217 are the galaxies with the
earlier Hubble type among our sample, it is possible that the position
of their hotspots in Figure~\ref{fig:SingleBurst} is dictated by a
larger intervening bulge population than in the other nuclear rings in
our sample, rather than by weaker star formation activity.

Although the properties of an SSP superimposed on an older bulge and
disc population can match well the hotspot line indices and provide
estimates for the age and mass fraction of the ring populations, in
practice such a scenario is quite implausible.
The young age of stars in the rings inferred from such models ($\la
30-100$\,Myr), combined with the relatively high fraction of spiral
galaxies with nuclear rings ($\sim20\%$, Knapen 2005), would imply
that we live in a special time in the history of spiral
galaxies. Instead, it is more likely that star formation in nuclear
rings has been either continous in the last few hundred Myr, or
characterised by multiple starbursting episodes.

\subsubsection{Continuous star formation in the ring}
\label{subsubsec:AbsLineDiagn_continuous}

To test the first of these classes of models, we compare
(Fig.~\ref{fig:ContinousBurst}) the position of the hotspots in the
[MgFe] vs. \Hb\ (upper panels) and D(4000) vs. \HdA\ diagrams (lower
panels) with the predicted evolution of these indices since the onset
of a period of constant star formation activity in the ring. As above,
this phase follows the main formation episode for the bulge and disc
stars by 10, 5, and 3\,Gyr (left, middle and right panels,
respectively). As in Figure~\ref{fig:SingleBurst}, the strength of the
star formation activity in the ring is a parameter that can be varied:
we show models which have added from 0.1\% to 10\% of the bulge and
disc mass after 100\,Myr.

There are several inconsistencies between the data and the models in
Figure~\ref{fig:ContinousBurst}. A continous star formation scenario
underpredicts the strength of the H$\delta$ absorption features in
many of the ring hotspot spectra, and cannot account for the values of
the \Hb\ index observed in NGC~7217. Furthermore, the position of
different hotspots within the same galaxies, like NGC~7742, NGC~5248
and NGC~4314, suggests that the different regions in the ring are
subject to different levels of constant star formation activity that
have begun at different epochs, several hundred Myrs in the past. This
is, however, a most unlikely situation, given that stars and gas
circulate along nuclear rings typically in a few tens of Myr.

\subsubsection{Multiple-burst star formation in the ring}
\label{subsubsec:AbsLineDiagn_multiple}

Discarding a history of constant star formation in the rings we now
consider the possibility of multiple starbursting episodes. In
Figure~\ref{fig:MultiBurst} we show the evolution of the line indices for
models including from 1 to 5 instantaneous bursts, each adding 1\% of
the bulge and disc mass and occurring every 100\,Myr. Because of this
periodicity, the model evolution is shown only for the 100\,Myrs
following the last star formation episode. Each of the models includes
a disc and bulge population that is 3\,Gyr old at the onset of the
latest burst, so that the age of the old stellar component is the same
for all models, whereas the recent episodic star formation activity
started earlier in models with more such episodes (e.g., star
formation in the ring began 400\,Myr ago, and 2.6\,Gyr after the
bulge/disk formation, in a 5-burst model).
Figure~\ref{fig:MultiBurst} also shows, for comparison, the composite
population models for an instantenous and continous star formation
scenario in the ring and a 3-Gyr-old disc and bulge population.

Except for NGC~4321 and for the rings requiring very small fractions
of young stars (NGC~4314 and NGC~7217), the hotspot data points on
both the [MgFe] vs. \Hb\ (upper panels) and D(4000) vs. \HdA\ diagrams
(lower panels) agree remarkably well with these models for multiple
starbursting episodes, suggesting that at least 4 or 3 bursts have
occurred in the past, as judged from the [MgFe] and D(4000) diagrams,
respectively.
Similarly to the case of the instantaneous models, we note that the
position of the hotspots in NGC~4314 and NGC~7217 in these diagrams
can also be matched by models featuring periodic star-formation
episodes, provided that each burst contributes to mass fractions
smaller than 1\% and that an older bulge and disc population is
included.

\subsubsection{Summary of star formation history modelling}

Despite the success of the last class of models, the star formation
history of nuclear rings is almost certainly more complicated than a
simple superposition of instantaneous bursts evenly spaced in time.
Some combination of the two latter scenarios is likely, in which star
formation events occur episodically, but may vary in length and
relative strength. At present, it is impossible to reproduce the {\it
exact\/} sequence of star formation in the rings from these
indices. What is apparent is that the models that most closely match
the data points must have a small, but significant, intermediate age
population that is the result of an extended period of recent star
formation. They are {\it not\/} the result of either continuous star
formation, or of one single burst of star formation (the current one).

\section{Emission Line Diagnostics: Physical Conditions of the
  Gas}\label{sec:EmiLineDiagn} 

In nuclear rings, photoionisation by hot OB stars is the most likely
source of nebular emission, although shocks may play an additional
role. A number of factors determine the appearance of the emission
line spectrum: the metallicity of the gas, the shape of the ionising
radiation spectrum, and the geometric distribution of the gas and
ionising sources (Dopita et al. 2000). The geometry can be combined
into one single factor, the mean ionisation parameter $q$ (with
dimensions cm\,s$^{-1}$). This is the ratio of the flux of ionising
photons through a unit area, to the local number density of hydrogen
atoms at the edge of the ionised cloud.  In this section, we inspect
the emission line spectra of our nuclear rings and nuclei, to
determine the gas density, the starburst metallicity, and the
ionisation parameter. To do this, we will measure the flux ratios of a
number of emission lines and compare the observed values to
photoionisation models.

\placefigeight
\placefignine

\subsection{Gas Density}
\label{subsec:EmiLineDiagn_density}

The two lines in the \Sii$\lambda\lambda 6716,6731$\AA\, doublet have
similar excitation levels, but different collisional de-excitation
rates. This results in the relative populations of the two levels
being mostly dependent on the gas electron density (Seaton 1954),
which can therefore be estimated from the intensity ratio of these
lines.
Figure~\ref{fig:SiiDensity} shows the
\Sii$\lambda6716$/\Sii$\lambda6731$ ratio for the nuclear rings in our
sample. The electron densities, indicated by the dashed lines, are for
a typical temperature of $T=10^4$\,K (Osterbrock 1989). The nuclear
rings show a wide range in electron density, both within each ring,
and across the sample. Most of the rings fall within the range
$N_{\mbox{e}}=$10-350\,cm$^{-3}$, with NGC~6951 and NGC~7217 having
higher densities, of up to 800\,cm$^{-3}$. This is consistent with
previous studies: Kennicutt, Keel \& Blaha (1989) found a wide range
of electron densities, from $N_{\mbox{e}}=$10-1000\,cm$^{-3}$, for a
number of H{\sc ii} regions, which included both disc and
circumnuclear H{\sc ii} regions. Similar results were found by Ho,
Filippenko \& Sargent (1997).

\subsection{Metallicity and Ionisation Parameter}
\label{subsec:EmiLineDiagn_Zandq}

Veilleux \& Osterbrock (1987; hereafter V087) devised a classification
system based on the ratios of forbidden to hydrogen lines in order to
separate gas emission originating in H{\sc ii} regions from nebular
activity powered by other sources of ionisation such as active
galactic nuclei.
Their ratios involve lines which are close together in wavelength,
minimising errors in calibration and reddening correction, such as the
\Oiii$\lambda$5007/\Hb\ ratio in blue spectra, and the
\Nii$\lambda$6583/H$\alpha$, [S{\sc ii}]$\lambda\lambda$6716,6731/\Ha\
and [O{\sc i}]$\lambda$6300/\Ha\ in red spectra. When the [O{\sc
iii}]/\Hb\ ratio is plotted against any of the red lines ratios,
different types of emission-line galaxies occupy different regions of
the diagrams. H{\sc ii} regions and H{\sc ii} nuclei in particular
tend to be located in narrow, well-defined zones (Kewley et al. 2001)
in such diagnostic diagrams, which can be understood by means of
sophisticated photoionisation models.

In this respect, Dopita et al. (2000) and Kewley et al. (2001) present
valuable models for H{\sc ii} regions obtained with the
photoionisation code MAPPINGS III (Sutherland \& Dopita (1993) while
using the PEGASE (Fioc \& Rocca-Volmerange 1997) and STARBURST99
(Leitherer et al. 1999) codes to generate the stellar
ionising-radiation fields that depend on the metallicity of the
starburst.
Figure~\ref{fig:EmiLinesDiagnostic} shows an example of a
photoionisation grid, calculated in the case of the \Oiii/\Hb\
vs. \Nii/\Ha\ diagnostic diagram for metallicities 0.05, 0.2, 0.4, 1,
and 2 times Solar, values of the ionisation parameter $q=3\times
10^8$, 1.5$\times 10^8$, 8$\times 10^7$, 4$\times 10^7$, 2$\times
10^7$, 1$\times 10^7$, and 5$\times 10^6$\,cm\,s$^{-1}$, and for
densities of $N_{\mbox{e}}=$10\,cm$^{-3}$ and 350\,cm$^{-3}$. The
narrow zone occupied by H{\sc ii}-regions and H{\sc ii}-nuclei in this
plot coincides with the folding of the ionisation
parameter/metallicity surface: H{\sc ii} regions characterised by a
large range in metallicity or ionisation parameter are projected into
a narrow band on the figure. Theses starburst models set tight upper
limits for the values of these line ratios, so that emission-line
regions that lie above and to the right of the model grids must have
other source of ionisation beside, or in addition to, O stars (Kewley
et al. 2001).

Although star-forming regions are readily identified in VO87 diagrams
such as the one juxtaposing [O{\sc iii}]/\Hb\ with \Nii/\Ha, the
folded nature of the models makes it difficult to determine
unambiguously the metallicity and the ionisation parameter of
star-forming regions, except for very low metallicities.
Dopita et al. (2000) and Kewley et al. (2001) found that the best
combination of line ratios for separating metallicity and ionisation
parameter effects are
\Nii$\lambda\lambda$6548,6583/\Oii$\lambda\lambda$3727,3729
vs. \Oiii$\lambda$5007/\Oii$\lambda\lambda$3727,3729, or \Nii/\Oii\
vs. \Oiii/\Hb.
Although all these lines fall within the wavelength range of our
spectra, we cannot measure the \Nii/\Oii\ ratio with confidence, since
we do not dispose of an accurate relative flux calibration of our red
and blue spectra, where the \Nii\ and \Oii\ lines are found,
respectively. Additionally, the \Oii\ lines are not always observed.
To estimate the metallicity and ionising parameter of the
emission-line regions observed by our spectra we can instead use the
diagnostic diagram based on the \Nii$\lambda$6583/\Sii\ and \Oiii/\Hb\
ratios. Rubin, Ford \& Whitmore (1984) were the first to suggest the
\Nii/\Sii\ ratio as a metallicity indicator, and Kewley \& Dopita
(2002) recently confirmed the usefulness of this ratio for estimating
metallicities from slightly sub-solar to super-solar values, making
the \Nii/\Sii\ quite complementary to \Nii/\Ha\ in this respect.
This is illustrated by the MAPPINGS III models plotted in the right
panel of Figure~9, where the model grid in the \Nii/\Sii\ diagram
folds only at very low metallicity.

Figure~9 shows the \Oiii/\Hb, \Nii/\Ha\ and \Nii/\Sii\ line ratios
measured in the ring hotspots. The location of the ring data in the
\Nii/\Ha\ diagram is consistent with nebular emission powered only by
O stars, although no conclusion on the metallicity of the starburst
can be drawn from this VO87 diagram.  On the other hand, the position
of the data points in the \Nii/\Sii\ vs. \Oiii/\Hb\ grid indicates
that the ring hotspots are populated by \Hii\ regions with
metallicities ranging from slightly below to just above Solar, with
values for the ionisation parameter between 1 and
2$\times10^{7}$\,cm\,s$^{-1}$.

Two aspects of this emission-line analysis are also relevant to the
our stellar population results.
First, Figure~\ref{fig:EmiLinesDiagnostic} firmly rejects starbursts
of exceedingly low metallicity, thus further excluding the possibility
discussed in \S~\ref{subsec:AbsLineDiagn_single} that the position of
some of the line-strength measurements in the ring hotspots could be
consistent with SSP models with subsolar metallicity.
Second, the models grid of Figure~\ref{fig:EmiLinesDiagnostic} use a
spectral energy distribution that was obtained from STARBURST99 and
assuming continuous star formation over a short period of
8\,Myr. Models based on a truly instantaneous star formation event
were found to be grossly inconsistent with the data, producing
exceedingly high \Oiii/\Hb\ ratio compared to the values measured in
the ring hotspots. This further supports the statement made in
\S~\ref{subsec:AbsLineDiagn_composite}, that the star-formation
history of nuclear rings is likely to be a combination of continous
and episodic bursts.

\subsection{Line ratios for the nuclei}
\label{subsec:EmiLineDiagn_nuclei}

All the observed nuclei in the sample have measurable emission
lines. Some of the galaxies are classified as AGN in the literature,
others as \Hii-nuclei or Low-Ionisation Nuclear Emission Regions
(LINERs, Heckman 1980). Baldwin, Phillips \& Terlevich (1981) and,
later, VO87 reported that the \Oiii/\Hb\ vs. \Nii/\Ha\ diagram allows
one to distinguish between the different types of nuclei. This work
has been extended by Kewley et al. (2001), who determined a
theoretical `maximum starburst line': galaxies above this line are
likely to be dominated by an AGN. Kauffman et al. (2003) modified this
scheme to include an empirically derived line dividing pure
star-forming galaxies from Seyfert-H{\sc ii} composite objects. These
classification lines can be seen in the top panel of
Fig.~\ref{fig:EmiLinesDiagnosticNucleus}, compared to the ratios
derived for the nuclear regions of the galaxies in the current
sample. NGC~5383 is closest to possessing a purely star-forming
nucleus, NGC~473 and NGC~5248 are classified as composite, and the
rest fall in the LINER and AGN category.

Kewley et al. (2006) have extended the scheme even further, using the
\Oiii/\Hb\ vs. \Sii/H$\alpha$ plot to separate AGN from LINER
activity. This can be seen for our data in the lower panel of
Figure~\ref{fig:EmiLinesDiagnosticNucleus}. In this diagram, NGC~6951
is the only Seyfert nucleus, NGC~7217 and NGC~7742 are well within the
region occupied by LINERs, NGC5383 and NGC5248 remain close to the
\Hii-region model grids, and each of the remaining nuclei lie on the
line between LINER and star-forming activity, suggesting a composite
nature. These results are in remarkable agreement with the Palomar
classification for these nuclei (Table~1), with the possible exception
of NGC4314, for which our data suggest a composite classification
rather than a strict LINER activity. NGC~473 was not observed by Ho et
al., and our data suggest a H2/T2 nuclear classification.

\section{Discussion}\label{sec:Discussion}

\subsection{The star-formation history of nuclear rings}
\label{subsec:Discussion_SFHrings}

In Paper~I we have investigated the star-formation history in the
nuclear ring of NGC~4321 using the three line-strength indices that
can be probed by the {\tt SAURON\/} integral-field spectrograph,
namely \Hb, Fe5015 and Mgb. For NGC~4321 we concluded that star
formation in the ring was likely long-lived and episodic, thus arguing
that the nuclear ring is a stable configuration that is intimately
linked to the large bar in this galaxy (Knapen et al. 1995b).

In the current paper we have measured the strength of two additional
stellar features, the 4000\AA\ break and the \Hd\ absorption line, in
the nuclear rings of eight nearby galaxies. By comparing the observed
values for all five indices to the predictions of population synthesis
models for the star-formation history in the ring we have reached a
similar conclusion to Paper~I - that star formation in nuclear rings
is a prolonged and episodic phenomenon.

Models assuming that nuclear rings form single instantaneous burst of
star formation yield predictions that are not incompatible with the
data, but such a scenario is generally ruled out because it implies
formation epochs for the rings that are too shot and recent to be
consistent with the relatively high frequency of occurrence of nuclear
rings in nearby galaxies
(\S~\ref{subsubsec:AbsLineDiagn_instantaneous}).
Furthermore, the inclusion of the D(4000) and \HdA\ indices in our
analysis allowed us to firmly rule out the possibility of a history of
continuous star formation for our nuclear rings
(\S~\ref{subsubsec:AbsLineDiagn_continuous}).
On the other hand, the strength of the absorption lines in the ring
hotspot spectra are well matched by models allowing for the formation
of the ring stars over a number of instantaneous star-formation
episodes, superimposed on an underlying or intervening old disc and
bulge population (\S~\ref{subsubsec:AbsLineDiagn_multiple}).

\placefigten

The success of this family of models, together with the demise of its
alternatives, confirm the general understanding that after the initial
formation of the bulge and disc component, gaseous material is
transported toward the center over a prolonged period of time, where
it accumulates near the location of the ILR and where the density
threshold for the ignition of star formation is repeatedly crossed.
The inflow of gaseous material can be driven by the presence of large
scale bars, but this is not the only mechanism that can fuel the
circumnuclear region (see, e.g., Knapen et al. 2004 for a case study
of ring formation induced by a minor merger). In fact, two of our
sample galaxies are classified as unbarred and they will be
further discussed below.
Another object that will deserve further attention is NGC~4314, for
which our modelling suggest that the ring stars have formed very
recently, only a few Myr ago. Below, we will discuss the possibility
that for NGC~4314 we could be witnessing the initial phases of ring
formation.

Our results on the star-formation history of nuclear rings have
several important consequences. As previously stated, nuclear rings
are common in spiral galaxies, and are characterised by high star
formation rates. If nuclear rings form stars over a long timescale,
this will lead to the the accumulation of considerable stellar masses
in the central regions of their host galaxies. Kormendy \& Kennicutt
(2004) estimated that for the typical mass of molecular gas and star
formation rate of nuclear rings, stellar masses of 10$^8$ to
10$^9$\,$M_{\sun}$ will be formed.
In Paper~I, the nuclear ring in NGC~4321 was observed to be thicker
when viewed in stellar indices than when viewed in \Hb\ emission,
indicating that the new stars are diffusing out of the ring. It is
possible that these stars will diffuse further in the galactic disc,
contributing to the formation of a more bulge-like structure due to
vertical instabilities triggered by the presence of a bar.
First characterised as pseudo-bulges by Kormendy (1993), such
structures half-way between bulges and discs may be an integral
component of most spiral galaxies, in particular in those of later
Hubble types. Kormendy \& Kennicutt (2004) further discuss how
pseudo-bulges are a consequence of slow secular processes, pointing
out how secular evolution becomes more important than galaxy mergers
or interactions in re-shaping the structure of galaxies as the
Universe expands.

If nuclear rings are stable configurations in disk galaxies that add
considerable stellar mass to the centers of their hosts over prolonged
periods of star formation, nuclear rings should be regarded as a key
component in the secular evolution of galaxies.
Furthermore, nuclear rings may be even more frequent than previously
thought. Most nuclear rings, and all those used by Knapen (2005) to
derive the fraction of nuclear ring in local spirals, were identified
using \Ha\ imaging that traces massive star formation. Yet, our
stellar population modelling suggest that over most of their lifetime
nuclear rings are not actively forming massive stars, which implies
that a significant but so far unknown fraction of galaxies could host
a young nuclear ring that is presently not forming massive stars.
The confirmation of nuclear rings as an established feature of a
galaxy also reinforces the view that nuclear rings are linked to the
large-scale bar, and the connection between nuclear rings and ILR(s).

\subsection{NGC~4314}
\label{subsec:Discussion_NGC4314}

Our stellar absorption-line measurements show how NGC~4314 stands
apart from the majority of the rings in our sample, with a
significantly older underlying or intervening disc and bulge
population and much younger ring stars.
This conclusion is consistent with the findings of Benedict et
al. (2002), who analysed high-spatial resolution optical colour maps
obtained with the Hubble Space Telescope (\HST) to study the ages of
the single star clusters in the ring. They found colours consistent
with instantaneous star formation and little evidence for any cluster
older than 40\,Myr in the ring.
Although the analysis of Benedict et al. did not cover the
intra-cluster stellar population, which also contribute to our hotspot
spectra, the absence of clusters older than 40\,Myr suggests that no
star-formation episode occured in the ring of NGC~4314 previous to the
current one.

It would thus appear that in the case of NGC~4314 we are witnessing
the initial massive star formation event, which may well coincide with
the formation of the nuclear ring.
The observed blue stellar arms just outside the nuclear ring (Benedict
et al. 2002) are evidence for star formation activity in the recent
past (between 100-200 Myr ago) but whether this occurred within the
present configuration of the ring, or within a larger past version of
it (as suggested by Combes et al. 1992), is not presently clear. What
is evident is that NGC~4314 is peculiar among our sample galaxies also
because the massive star formation in the entire galaxy is practically
confined to the nuclear ring (known since Burbidge \& Burbidge 1962),
and because the galaxy is particularly deficient of atomic hydrogen
gas, H{\sc i} (Garc\'\i a-Barreto et al. 1991; Combes et al. 1992).
These peculariaties make the conclusion that we are indeed witnessing
the very first phase of massive star formation in the nuclear ring
less unattractive than it might seem at first glance when compared to
the results for the other rings in our sample.

\subsection{Unbarred galaxies: NGC~7217 and NGC~7742}
\label{subsec:Discussion_unbarred}

NGC~7217 and NGC~7742 are both classified as unbarred galaxies and
indeed appear very symmetric in near-IR 2MASS images.
Using ground-based and \HST\ near-IR imaging, Laine et al. (2002)
confirm that no bars can be found in NGC~7217, but find two small bars
in NGC~7742 (with deprojected bar radii of 1.2 and 6.8 arcsec,
corresponding to 130 and 730 parsec).
The lack of a large bar in both cases may initially seem
problematic, as the general picture of nuclear ring formation is often
described as requiring a bar to drive gas towards the centre. There
are other ways of achieving a non-axisymmetric potential which can
stimulate inflow, however, such as the presence of an oval distortion,
or the influence of external forces. Both NGC~7217 and NGC~7742 have
counter-rotating sub-systems (Merrifield \& Kuijken 1994; Sil'chenko
\& Afanasiev 2000; de Zeeuw et al. 2002), which hints at a turbulent,
merger, history.

NGC~7742 has a counter-rotating gas system, and NGC~7217 has a
counter-rotating stellar system. Sil'chenko \& Moiseev (2006) suggest
that the two galaxies have undergone a similar past interaction. They
propose that NGC~7742 has only recently started accreting gas, while
NGC~7217 is at a later stage of evolution and has already converted
its counter-rotating gas into stars.
Mazzuca et al. (2006) find that the metallicity of the gas in the ring
of NGC~7742 is close to Solar, in agreement with our results.
As there are no suitable large metal-rich companion galaxies with
which NGC~7742 can have interacted, Mazzuca et al. propose that a
smaller, metal-poor dwarf galaxy has been cannibalised. Part of its
gas was then concentrated in the ring, where star formation proceeded
long enough for the metallicity to reach a solar value. They estimate
that this process takes a few tens of Myrs, which agrees with our
stellar population modelling for this galaxy.

Buta et al. (1995) and Combes et al. (2004) suggest that the oval
distortion present in NGC~7217 would be sufficient to create the
resonances which shape the three rings which it possesses, without
invoking a minor merger explanation, and the detection of a
counter-rotating stellar system could be attributed to a large bulge
velocity dispersion. They also highlight that a number of nuclear
rings is present in NGC~7217, all found at different radii. The
smallest is a red dust ring, followed radially outwards by an \Ha\
ring, in turn followed by a broad red ring.
Like NGC~4314, NGC~7217 is also deficient in \Hi\ gas (Buta et
al. 1995), and both have much older underlying stellar populations
(Fig.~\ref{fig:IndexIndexDiagrams}) than the other galaxies in our
sample.

\subsection{Absence of Wolf-Rayet features}
\label{subsec:Discussion_WR}

Wolf-Rayet stars (WR) are the remnants of the most massive stars in
galaxies. Their numbers therefore peak shortly after a major
star-formation episode, typically when the starburst is only 3 to
4-Myr-old. The number and type of WR stars is also a strong function
of the starburst metallicity, a fact that stimulated the study of WR
stars as a means to trace the nature of massive star formation in
external galaxies (e.g., Vacca \& Conti 1992; Shaerer \& Vacca 1998;
Shaerer, Contini \& Kunth 1999; Guseva, Izotov \& Thuan 2000). In
optical spectra, the most noticeable WR feature is the so-called
4650\AA\ ``bump'', which is due to a superposition of broad stellar
emission lines.

In {\it none\/} of our ring hotspot spectra is such a feature
observed.  The absence of WR features suggests that the {\it bulk\/}
of the young stars in the ring regions subtended by our spectra have
formed over a longer period of time than just a few Myrs, consistent
with our absorption-line analysis.
The only exception to this statement is NGC~4314. For this galaxy we
have evidence that the hotspot spectra are a superposition of an old
bulge and disc stellar population with very young ring stars, less
than 25-Myr-old. A 4650\AA\ ``bump'' could therefore be expected in
this case, and that this feature is not observed could be due to a
strong contamination of intervening bulge stars, consistent with the
early-type classification of this disc galaxy (see also
\S~\ref{subsubsec:AbsLineDiagn_instantaneous})

\section{Conclusions}\label{sec:Conclusions}

We present a study of the optical spectra of a sample of eight
star-forming nuclear rings and the nuclei of their host
galaxies. These spectra cover a wide range in wavelength, and allow us
to measure a number of gaseous and stellar features. The stellar
absorption indices show that along the line of sight to the nuclear
ring we can detect the older underlying and intervening disc and
bulge, as well as the young stars being formed in the ring.
We use two absorption-line index diagrams, \Hb\ vs. MgFe and \HdA\
vs. D(4000), to test models for the star formation history of the
rings. The simultaneous use of these indices allows us to rule out a
constant star-formation history, whereas the possibility that these
rings one-off events in a galaxy's history is discarded on the basis
that it would be unlikely to observe all these rings as they have just
formed.
Our models suggest instead that nuclear rings are in fact more
long-lived structures with extended, although episodic, periods of
star formation. The absence of Wolf-Rayet features in our spectra
further shows that nuclear rings are not made exclusively of stars
that formed only very recently.
These results confirm not only the intimate dynamical connection
between nuclear rings and their host galaxies, but also the important
role that the rings can play in transforming disk gas to
(pseudo-)bulge stars, thus stimulating secular evolution.

The gas emission lines of the nuclear rings reveal that the rings
possess a range of electron densities, most within a range of
10-350\,cm$^{-3}$. Diagnostic diagrams show that all rings are
populated by \Hii-regions with similar ionisation parameters and with
a small spread in metallicity around the Solar value.

The nuclear rings of NGC~4314 and NGC~7217 stand apart from the other
nuclear rings in the sample, as they appear to have much older and
prominent underlying populations. They also have much lower values of
the \Hb\ and \HdA\ indices, which in the case of NGC~4314 implies much
younger ages for the ring stars.
In this galaxy, which stands out from the rest of the sample for its
lack of star formation in the disk and its deficiency of neutral gas,
we may be witnessing the initial phases of the ring formation.
The vast majority of the nuclear rings in the sample can thus be
described by a scenario where the nuclear ring is a stable structure,
in most cases induced by the bar, and confirming that all these rings
have formed by dynamical means near the ILR(s). This is also true for
the two galaxies in our sample which lack a prominent large bar, but
for which a past minor merger event may well have caused a similar
non-axisymmetry in the galaxy's gravitational potential.

Our study of the spectra of the nuclei of our sample galaxies shows
that all display emission lines. Consistent with previous studies for
the nuclear activity and stellar populations, the majority of our
nuclei appear to be dominated by old stellar populations and by
LINER-like emission.

\section*{Acknowledgments}

We thank the referee, Herv\'e Wozniak, for his useful comments. This
work is based on observations made with the WHT, operated on the
island of La Palma by the Isaac Newton Group in the Spanish
Observatorio del Roque de los Muchachos of the Instituto de
Astrof\'{i}sica de Canarias. ELA was supported by a PPARC
studentship. JHK acknowledges the Leverhulme Trust for the award of a
Leverhulme Research Fellowship.

\bsp

\label{lastpage}

\end{document}